\documentclass[english]{article}
\usepackage[T1]{fontenc}
\usepackage[utf8]{inputenc}
\usepackage{verbatim}
\usepackage{amsmath}
\usepackage{amsthm}
\usepackage{amssymb}
\usepackage{stmaryrd}
\usepackage{color}
\usepackage{graphicx}
\usepackage{subfig}
\usepackage{float}
\usepackage{enumitem}

\usepackage{hyperref}
\hypersetup{
   colorlinks=true,%
   citecolor=black,%
   filecolor=black,%
   linkcolor=black,%
   urlcolor=blue
}

\usepackage{tikz}
\newcommand*\circled[1]{\tikz[baseline=(char.base)]{
             \node[shape=circle,draw,inner sep=.8pt] (char) {#1};}}

\makeatletter
\theoremstyle{plain}
\newtheorem{thm}{\protect\theoremname}
  \theoremstyle{plain}
  \newtheorem{lem}[thm]{\protect\lemmaname}
  \theoremstyle{plain}
  \newtheorem{prop}[thm]{\protect\propositionname}
  \theoremstyle{definition}
  \newtheorem{defn}[thm]{\protect\definitionname}
\newtheorem{example}[thm]{Example}
\newtheorem{rem}[thm]{Remark}
\newtheorem{cor}[thm]{Corollary}

\def \I {\mathbb{I}}

\def \l {\left}
\def \r {\right}

\def \oline {\overline{g}}
\def \pline {1}

\def \interior {\operatorname{interior}}

\def \der{\mathrm{d}}
\def \d{\delta}
\def \F{\mathrm{F}}
\def \W{\mathrm{W}}
\def \G{\mathrm{GW}}

\makeatother

\usepackage{babel}
  \providecommand{\definitionname}{Definition}
  \providecommand{\lemmaname}{Lemma}
  \providecommand{\propositionname}{Proposition}
\providecommand{\theoremname}{Theorem}

\newcommand{\editr}[1]{{\color{black} #1}} % Revision
\newcommand{\edita}[1]{{\color{black} #1}}

\newcommand{\edit}[1]{{\color{black} #1}}

\begin{document}

\title{Stable Super-Resolution of Images: A Theoretical Study}

\author{Armin Eftekhari, Tamir Bendory, and Gongguo Tang\thanks{AE is with the Department of Mathematics and Mathematical Statistics at Umea University. TB is with   
% the School of Electrical Engineering at Tel Aviv University. 
the Program
in Applied and Computational Mathematics at Princeton University. 
GT is with the Department of Electrical Engineering at the Colorado School of Mines.}}
\maketitle

\begin{abstract}

We study the ubiquitous \edit{{super-resolution} problem}, in which one aims at localizing {positive} point sources in an image, blurred by the  point spread function of the imaging device. To recover the point sources, we  propose to solve a convex feasibility program, which simply  finds {a} nonnegative Borel measure that agrees with the observations collected by the imaging device.

 In the absence of imaging noise, we show that solving this convex program uniquely retrieves the point sources, {provided that the} imaging device collects enough observations. This result holds true if the point spread function \edit{of the imaging device} can be decomposed into horizontal and vertical components, and if the translations of these components  form a Chebyshev system, i.e., a system of continuous functions that loosely behave like algebraic polynomials.

Building {upon} recent results for one-dimensional signals~\cite{eftekhari2018sparse}, 
we  prove that this super-resolution algorithm is stable\editr{, in the generalized Wasserstein  metric,} to  model mismatch (i.e., when the image is not  sparse) and to additive imaging noise. In particular,  the 
recovery error depends on the noise level and how well the image can be approximated with well-separated point sources. As an example, we verify these claims for the important case {of  a Gaussian point spread function}. The proofs rely on the construction of novel interpolating polynomials\editr{---which are the main technical contribution of this paper---}\edit{and partially resolve the  question raised in \cite{schiebinger2015superresolution} about the extension of the standard machinery to higher dimensions. } 
\end{abstract}

\section{Introduction \label{sec:intro}}

Consider an unknown number of point sources with unknown locations and amplitudes. An imaging mechanism provides us with a few noisy measurements from which we wish to estimate the locations and amplitudes of these sources. Because of the finite {resolution} of \editr{any} imaging device, poorly separated sources are indistinguishable without using \editr{an appropriate} localization \editr{technique} that \edit{would take} into account the \editr{sparse structure within the image}.

This {super-resolution} problem of localizing point sources finds various applications in, \editr{for instance}, astronomy \cite{puschmann2005super},  geophysics \cite{khaidukov2004diffraction}, chemistry, medicine, microscopy and neuroscience \cite{betzig2006imaging,hess2006ultra,rust2006sub,ekanadham2011neural,hell2009primer,tur2011innovation,solomon2017sparcom}. 
% Most of these applications involve images or higher dimensional signals.  
In this paper, we study the grid-free and \editr{nonnegative} super-resolution of two-dimensional {(2-D)} signals \edit{(i.e., images)} in the presence of noise, extending the one-dimensional {(1-D)} results of~\cite{eftekhari2018sparse}.

Let $x$ be a {nonnegative} Borel measure supported on $\I^2=[0,1]\times [0,1]$\edit{, and let $\{\phi_{m}\}_{m=1}^M$ be real-valued and continuous functions.
\editr{We model} the \edit{(possibly noisy)}  observations
 $\{y_{m,n}\}_{m,n=1}^M$ collected from $x$ as} 
\begin{equation}
y_{m,n}\approx \int_{\I^{2}}\phi_{m}(t)\phi_{n}(s)\,x\left(dt,ds\right).
\label{eq:approximate}
\end{equation}
More specifically, we assume that 
\begin{equation}
\sum_{m,n=1}^M\left|y_{m,n}-\int_{\I^{2}}\phi_{m}(t)\phi_{n}(s)\,x\left(dt,ds\right)\right|^{2}\le\delta^{2},\label{eq:noise}
\end{equation}
where $\delta\ge0$ reflects the additive noise level. {We do not impose \editr{a} statistical model \editr{for} the noise.}  If we define the matrices
$y\in\mathbb{R}^{M\times M}$ and $\Phi(t,s)\in\mathbb{R}^{M\times M}$
such that 
\begin{equation}
y[m,n]=y_{m,n},\qquad\Phi(t,s)[m,n]=\phi_{m}(t)\phi_{n}(s),
\qquad \forall m,n\in [M]:=\{1,\cdots,M\},
\label{eq:shorthands}
\end{equation}
we may rewrite (\ref{eq:noise}) more compactly   as 
\begin{equation}
\l\|y-\int_{\I^{2}}\Phi(t,s)\,x(dt,ds)\r\|_{\mathrm{F}}\le\delta,
\label{eq:meas model}
\end{equation}
where $\|\cdot\|_{\mathrm{F}}$ stands for the Frobenius norm. \edit{Often, $\phi_m$ and $\phi_n$ above are translated copies of a function $\phi$,}   $\phi(t)\phi(s)$ is {referred to as the} {point spread function} of the imaging device, and $y$ is {the {2-D} acquired signal that can be thought of as an image} with $M^2$ pixels. \edit{We note that the {tensor product} model in \eqref{eq:meas model} is widely used as a model in imaging \cite{Candes2014,bendory2017robust,morgenshtern2014stable}.}   For example, if the imaging device acts as an ideal low-pass filter with the cut-off frequency of $f_c$, \edit{then the corresponding choice is} $\{\phi_m\}_{m=1}^M = \{\cos(2\pi kt)\}_{k=0}^{f_c}\cup \{\sin(2\pi k t)\}_{k=1}^{f_c}$ with $M=2f_c+1$. \edita{It is also possible to collect observations using two different set of functions along $t$ and $s$ directions ($\{\phi_m(t)\}_m$ and $\{\psi_n(s)\}_n$). \editr{However, for the sake of clarity}, we avoid this additional layer of complexity here.}

{In order to recover $x$, we suggest using the simple convex feasibility program}
%We attempt to recover
%$x$ by solving the convex feasibility program 
\begin{equation}
\text{find a nonnegative Borel measure }z\text{ on }\I^2 \text{ such that }\l\|y-\int_{\I^{2}}\Phi(t,s)\,z(dt,ds)\r\|_{\mathrm{F}}\le\delta',\label{eq:feas}
\end{equation}
\edit{for some $\delta'\geq\delta$, which is reminiscent of nonnegative least squares in finite dimensions \cite{slawski2013non,foucart2014sparse}.} \edita{Once Program~\eqref{eq:feas} is solved, the zeros of the optimal dual function can be used as \editr{estimates for the} locations of the point sources. Alternatively, one may apply the Prony's method~\cite{weiss1963prony}  or the matrix pencil approach~\cite{hua1990matrix} to a solution of Program~\eqref{eq:feas}, which is a measure on $\mathbb{I}^2$, to locate the point sources.}

\edit{ Program~\eqref{eq:feas} does not involve a grid on $\I^2$, and notably does {not} regularize $z$ beyond nonnegativity, thus radically deviating from the existing literature \cite{morgenshtern2014stable,bendory2017robust,schiebinger2015superresolution,de2012exact,denoyelle2017support}.
} This paper establishes that in the noiseless setting $\delta=0$, solving Program~\eqref{eq:feas} \edit{precisely} recovers the true  measure $x$, \edit{provided that} $x$ is a nonnegative sparse measure on $\I^2$ and under \edit{certain} conditions on the imaging apparatus $\Phi$. \editr{In addition}, when $\delta >0$ and $x$ is an arbitrary nonnegative measure on $\I^2$, solving Program~\eqref{eq:feas} well-approximates $x$. In particular, we establish that \emph{any} nonnegative measure supported on $\I^2$ that agrees with the observations $y$ {in the sense of \edit{Program}~\eqref{eq:feas} is near the true measure $x$}. % in generelazied Wasserstein metric.

{This paper does not focus on the important question of how to \edit{numerically} solve the infinite-dimensional Program~\eqref{eq:feas} in practice. One straightforward approach would be to discretize} the measure $z$ on a fine uniform grid for $\I^2$, thereby replacing Program~\eqref{eq:feas} with a finite-dimensional convex feasibility program that can be solved with standard convex solvers. 
\editr{Moreover,} a few recent papers \editr{have} proposed algorithms to \edit{directly} solve Program~\eqref{eq:feas}~\cite{eftekhariBridge,boyd2017alternating,bredies2013inverse}, \edita{i.e., these  algorithms can  be used to solve Program~\eqref{eq:feas} without discretization.} 
% \editr{Finally}, there are other popular techniques to estimate sparse measures, for instance, using various generalizations of Prony's method; see Section~\ref{sec:related} \edit{for a survey of the related literature}. 
A comprehensive numerical comparison between \editr{these} alternatives 
% for different noise levels 
is of great importance and we leave that to a future study.
This paper \editr{instead} aims to provide theoretical justifications for the success of Program~\eqref{eq:feas}, thereby arguing that \edit{imposing nonnegativity is \editr{theoretically} enough for successful super-resolution}. In other words,  under \edit{mild} conditions, the imaging device  acts as an injective map on sparse nonnegative measures {and we can stably find its inverse \editr{map}.}

This work  relies heavily on a recent work~\cite{eftekhari2018sparse}, which established that grid-free and \editr{nonnegative} super-resolution in {1-D} can be  achieved by solving the 1-D version of Program~\eqref{eq:feas}. {In doing so,} it removed the regularization required in prior work and substantially simplified the existing results. 
\editr{However,} extending~\cite{eftekhari2018sparse} to two dimensions is far from  trivial {and requires a careful design of \editr{a} new family of dual certificates, as will become clear in the next sections.} \edit{Indeed, this work overcomes the technical obstacles noted in   \cite[Section 4]{schiebinger2015superresolution} for extending the proof machinery to higher dimensions.}

Before turning to the details, let us summarize the technical contributions of this paper.
{Section~\ref{sec:main} presents these \editr{contributions} in detail\edit{, while} \editr{proofs} are deferred to Section~\ref{sec:theory} and the appendices.}

\paragraph{Sparse measures without noise. }
Suppose that \editr{the measure} $x$ consists of $K$ positive impulses located in $\I^2$. In the absence of noise ($\delta=0$), Proposition \ref{prop:noisefree} below shows that solving Program \eqref{eq:feas} with $\delta'=0$ successfully recovers $x$ from the observations $y\in\mathbb{R}^{M\times M}$, provided that $M\ge 2K+1$ and that  $\{\phi_m\}_{m=1}^M$  form a {Chebyshev system} on $\I$. A Chebyshev system, or $\mathcal{C}$-system for short,\footnote{\editr{It is also common to use T-system as the abbreviation of the Chebyshev system.}}  is a collection of continuous functions that loosely behave like algebraic monomials\edit{;} see Definition \ref{def:(T-systems)-Real-valued-and}. $\mathcal{C}$-system is a widely-used concept in classical approximation theory \cite{karlin1966tchebycheff,karlin1968total,kreinMarkov} 
{that also plays a {pivotal} role in some  modern signal processing applications}{; see for instance} 
\cite{eftekhari2018sparse,de2012exact,schiebinger2015superresolution}. In other words, Proposition~\ref{prop:noisefree} below  establishes that the imaging operator $\Phi$ in~\eqref{eq:meas model} is an injective map from $K$-sparse nonnegative  measures on $\I^2$, provided that $\{\phi_m\}_{m=1}^M$ form a $\mathcal{C}$-system on~$\I$ \edit{ and $M\ge 2K+1$}.   

{In contrast to earlier results,} no minimum separation between the impulses is necessary, {Program~\eqref{eq:feas} does not contain any explicit regularization  to \edit{promote} sparsity}, and \edit{lastly} $\{\phi_m\}_{m=1}^M$  need only to be continuous.  We note that Proposition~\ref{prop:noisefree} is a nontrivial extension of the 1-D result in~\cite{eftekhari2018sparse} to images. Indeed, the key concept of $\mathcal{C}$-systems do not generalize to two or higher dimensions and proving Proposition~\ref{prop:noisefree} requires a \edit{novel} construction of  dual certificates \edit{to overcome the technical obstacles anticipated in  \cite[Section 4]{schiebinger2015superresolution}. 
% see also the  general results below.
}

\paragraph{Arbitrary measure with noise.}
More generally, consider an arbitrary nonnegative measure $x$ supported on $\I^2$. As detailed \edit{later}, \edit{given} $\varepsilon\in (0,1/2]$, \editr{the measure} $x$ can always be approximated with a $K$-sparse and $\varepsilon$-separated  nonnegative measure, up to an  error of $R(x,K,\varepsilon)$ in the generalized Wasserstein metric, \editr{denoted throughout by $d_{\G}$}. {This is true} even if $x$ itself is not $\varepsilon$-separated or not atomic at all. We may think of  $R(x,K,\varepsilon)$ as the ``model-mismatch'' of approximating $x$ with a well-separated sparse measure, \edita{i.e., $R(x,K,\varepsilon) = d_{\G}(x,x_{K,\varepsilon})= \min d_{\G}(x,\chi)$, where the minimum is taken over every nonnegative, $K$-sparse and $\varepsilon$-separated measure $\chi$. }

In the presence of noise and numerical inaccuracies ($\delta\ge0$), Theorem~\ref{thm:main noisy simplified} below shows that solving Program~\eqref{eq:feas}  approximately recovers $x$ from the observations $y\in\mathbb{R}^{M\times M}$ in  the generalized Wasserstein metric $d_{GW}$. In particular, a solution $\widehat{x}$ of Program~\eqref{eq:feas} satisfies 
\begin{equation}
d_{\G}(x,\widehat{x}) \le c_1 \delta + \editr{c_2(\varepsilon)}+ c_3  R(x,K,\varepsilon),
\label{eq:demo}
\end{equation}
provided that $M\ge 2K+2$\edit{,} and  the imaging apparatus and certain  functions forms a $\mathcal{C}^*$-system, a natural generalization of the $\mathcal{C}$-system \editr{introduced earlier}. 
The  factors $c_1,c_2,c_3$ above are specified in the proof and depend chiefly on the measurement functions $\{\phi_m\}_{m=1}^M$, \edit{see \eqref{eq:shorthands}}.  Note that the recovery error in~\eqref{eq:demo} depends on the noise level $\delta$, the separation $\varepsilon$, and on how well $x$ can be approximated with a $K$-sparse and $\varepsilon$-separated measure, similar to the 1-D results in~\cite{eftekhari2018sparse}. 
\editr{In particular, as we will see later,} when $\delta = \varepsilon
= R(x,K,\varepsilon) =0$, \eqref{eq:demo} reads \editr{as} $d_{\G}(x,\widehat{x}) = 0$, and Theorem~\ref{thm:main noisy simplified} reduces to Proposition~\ref{prop:noisefree} for sparse and noise-free super-resolution. 

\edit{We remark} that Theorem \ref{thm:main noisy simplified} applies to any nonnegative measure $x$, without requiring any separation between the impulses in $x$. In fact, $x$ might not be atomic at all. Of course, the recovery error $d_{\G}(x,\widehat{x})$ does depend on how well $x$ can be approximated with a well-separated sparse measure, which is reflected in the right-hand side of \eqref{eq:demo} and hidden in the factors $c_1,c_2,c_3$ therein.  As emphasized earlier, no regularization other than nonnegativity \editr{is} used and $\{\phi_m\}_m$ need only be continuous.

As a concrete example of this general framework, we consider the case where $\{\phi_m\}_m$ are translated \edita{copies of a Gaussian ``window'', i.e., copies of a Gaussian function}. Building on the results from~\cite{eftekhari2018sparse}, we show in  Section~\ref{sec:gaussian} that  the conditions for
both Proposition~\ref{prop:noisefree} and Theorem~\ref{thm:main noisy simplified} are {met} {for this important example. That is, solving Program~\eqref{eq:feas} successfully and stably recovers an image that has undergone Gaussian blurring.}

\section{Main Results}\label{sec:main}

\subsection{Sparse Measure Without Noise}\label{sec:noisefree}

{Let $x$ be \editr{the} nonnegative atomic measure  
	\begin{equation}
	x=\sum_{k=1}^K a_{k}\cdot\delta_{\theta_{k}}, \quad a_k>0,
	\label{eq:atomic}
	\end{equation}
	with $K$ impulses located at $\Theta=\{\theta_{k}\}_{k=1}^K\subset\text{interior}(\I^{2})$
	\editr{and} positive amplitudes $\{a_{k}\}_{k=1}^K$. Here, $\delta_{\theta_{k}}$
is the Dirac measure located at $\theta_{k}=(t_{k},s_{k})$.
We first consider the case where there is no imaging noise ($\delta=0$), and thus we collect the noise-free observations 
}
\begin{equation}
y=\int_{\I^2} \Phi(\theta)\, x(d\theta)\in \mathbb{R}^{M\times M}.
\label{eq:noiseFreeMeas}
\end{equation}
To {understand} when solving Program~\eqref{eq:feas} with $\delta'=0$ successfully recovers the true measure $x$, recall the concept of $\mathcal{C}$-system~\cite{karlin1966tchebycheff}:
\begin{defn}[$\mathcal{C}$-system]
\emph{\label{def:(T-systems)-Real-valued-and}Real-valued
and continuous functions $\{\phi_{m}\}_{m=1}^M$ form a $\mathcal{C}$-system
on the interval $\I$, \editr{provided that} the \editr{determinant of the} $M\times M$ matrix $[\phi_{m}(\tau_{k})]_{k,m=1}^M$
is {\color{black}positive} for every {\color{black}(strictly)} increasing sequence $\{\tau_{k}\}_{k=1}^M\subset \I$. }
\end{defn}

\noindent For example, the monomials $\{1,t,\cdots,t^{M-1}\}$ form a $\mathcal{C}$-system on any closed interval of the real line. \edit{In fact, $\mathcal{C}$-system can be \editr{interpreted} as a generalization of ordinary monomials.} For instance, it is not difficult to verify that any  ``polynomial'' $\sum_{m=1}^M b_m \phi_m(t)$ of a $\mathcal{C}$-system $\{\phi_m\}_{m=1}^M$ has at most $M-1$ distinct zeros on \editr{the interval} $\I$. Or,  given $M$ distinct points, there exists a unique polynomial of $\{\phi_m\}_{m=1}^M$ that interpolates these points. Note also that \edit{the} linear independence of $\{\phi_m\}_{m=1}^M$ is a necessary{---but not sufficient---}condition for forming a $\mathcal{C}$-system. 
\editr{As an example} in the context of super-resolution, translated copies of the Gaussian window $e^{-t^2}$  form a $\mathcal{C}$-system on any interval \editr{of the real line}, and so do many other  windows~\cite{karlin1966tchebycheff}. \edit{As we will see later, the notion of $\mathcal{C}$-system allows us to design  a nonnegative polynomial  with prescribed zeros on \editr{the interval} $\I$, \editr{and this polynomial will play} a key role in} \editr{establishing the main results of this paper.} %\editr{establishing the following result}.}
%, which is in turn the  building  block for our construction of the dual certificate of Program \eqref{eq:feas} to prove Proposition~\ref{prop:noisefree} below.}

Proved in Section \ref{sec:nonoise proof}, \editr{the following result} states that solving Program~\eqref{eq:feas} successfully recovers $x$ from the noise-free image $y$, provided that the measurement functions form a $\mathcal{C}$-system.
\begin{prop}[Sparse measure without noise]
\label{prop:noisefree}Let $x$ be a $K$-sparse nonnegative   measure supported on $\operatorname{interior}(\I^2)$, \editr{specified by} \eqref{eq:atomic}. \edit{\editr{Suppose that} $M\ge 2K+1$ and}  that the measurement functions 
$\{\phi_{m}\}_{m=1}^M$ form a $\mathcal{C}$-system on \editr{the interval} $\I$.  \edit{Lastly,} \editr{for} $\delta=0$, consider the imaging operator $\Phi$ and the image $y\in\mathbb{R}^{M\times M}$ in \eqref{eq:shorthands}~and~\eqref{eq:meas model}. 
Then, $x$ is the unique solution of Program~\eqref{eq:feas} with $\delta'=0$. 
\end{prop}
In words, Program \eqref{eq:feas} successfully localizes the $K$ impulses present in \editr{the measure} $x$ \editr{from}  $(2K+1)^2$ measurements, \editr{provided that} the measurement functions $\{\phi_m\}_{m=1}^M$ form a $\mathcal{C}$-system on \editr{the interval} $\I$. {Note that {no} minimum separation is required between the impulses, in contrast to similar results for super-resolution with \edit{both} signed \edit{and nonnegative} measures; see for instance~\cite{Candes2013,bendory2016robust,bendory2017robust}. In addition, 
no regularization was imposed  in Program~\eqref{eq:feas} beyond nonnegativity,  and the measurement functions only need to be continuous.} 

\begin{figure}[ht!]
\begin{center}
\subfloat[\label{fig:impulseLocations}]{\protect\includegraphics[width=0.30\textwidth]{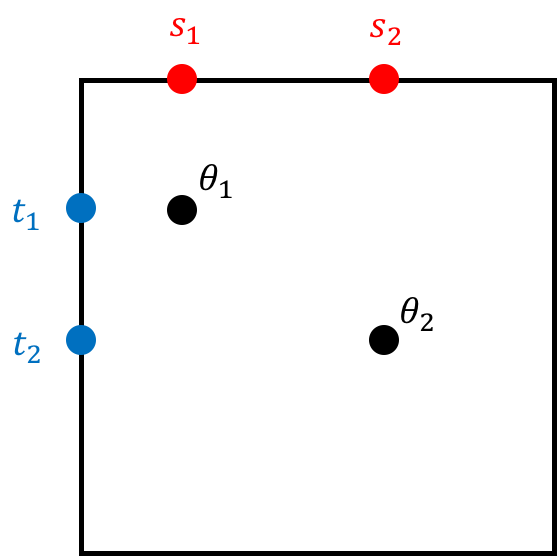}
}

\subfloat[\label{fig:polynomial1}]{\protect\includegraphics[width=0.30\textwidth]{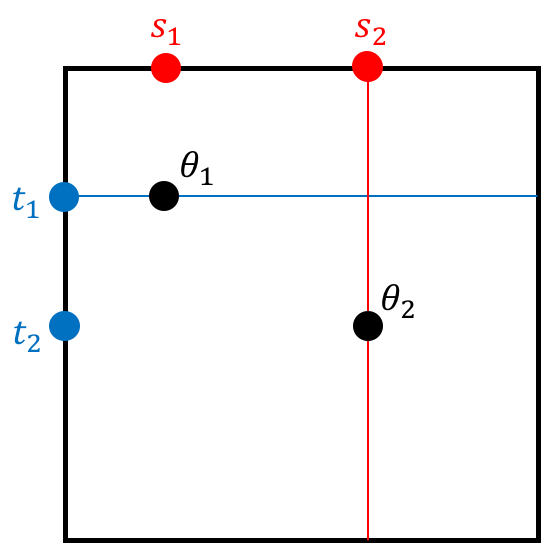}
}
\subfloat[\label{fig:polynomial2}]{\protect\includegraphics[width=0.30\textwidth]{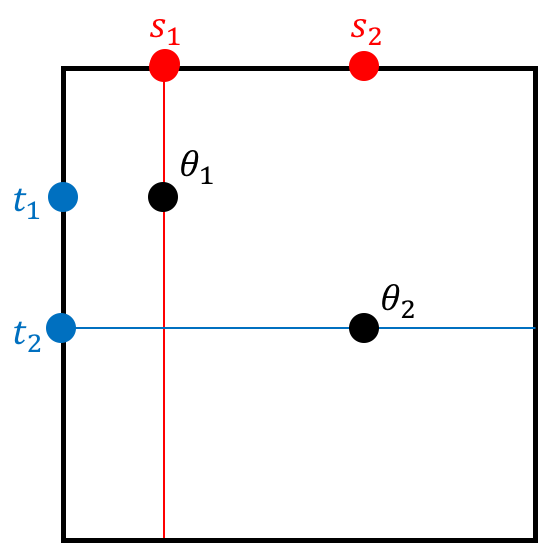}
}

\caption{This figure \editr{visualizes} the idea behind the proof of Proposition \ref{prop:noisefree}, \editr{explained in Remark~\ref{rem:proofTechNoiseFree}}.   As an example, consider the nonnegative measure $x=a_1\delta_{\theta_1}+a_2\delta_{\theta_2}$. The locations of two impulses at $\theta_1=(t_1,s_1)\in \I^2$ and $\theta_2=(t_2,s_2)\in \I^2$ are shown with black dots in Figure \ref{fig:impulseLocations}. For Program \eqref{eq:feas} to successfully recover $x$ from the image $y$ in \eqref{eq:noiseFreeMeas}, we should construct a nonnegative polynomial $Q$ of the form \editr{of}~\eqref{eq:form of dual polynomial} that has zeros exactly on $\theta_1$ and $\theta_2$. We do so by combining a number of univariate polynomials in $t$ and $s$. More specifically, consider a nonnegative polynomial $q_{t_1}(t)=\sum_{m=1}^M b^{11}_m \phi_m(t)$ that vanishes only at $t_1$. Likewise, consider similar nonnegative polynomials $q_{t_2}(t)$, $q_{s_1}(s)$, and $q_{s_2}(s)$, which are zero only at  $t_2$, $s_1$, and $s_2$, respectively. Figure~\ref{fig:polynomial1} shows the zero set of the polynomial {$q_{t_1}(t)q_{s_2}(s)$} as the union of blue and red lines. 
Similarly, Figure~\ref{fig:polynomial2} shows the zero set of the polynomial $q_{t_2}(t)q_{s_1}(s)$. Note that the intersection of these two zero sets is exactly $\{\theta_1,\theta_2\}$. That is, $q(\theta) = q_{t_1}(t)q_{s_2}(s)+q_{t_2}(t)q_{s_1}(s)$ is a nonnegative polynomial of the form in~\eqref{eq:form of dual polynomial} that has zeros only at $\{\theta_1,\theta_2\}$, as desired. It only remains now to construct the univariate polynomials $q_{t_1},q_{t_2},q_{s_1},q_{s_2}$ described above. When the \editr{measurement functions} $\{\phi_m\}_{m=1}^M$ form a $\mathcal{C}$-system and {$M\geq 2K+1$}, the existence of these univariate polynomials follows {from \editr{standard results}}. We {note} that the construction of $Q$ in this {example} is slightly different from the proof of Proposition~\ref{prop:noisefree}, \editr{in order} to simplify \editr{this} presentation. \label{fig:idea of proof noise free}} 
\end{center}
\end{figure}

\begin{rem}[Proof technique]\label{rem:proofTechNoiseFree} Let us outline the proof of Proposition \ref{prop:noisefree}. Loosely speaking, a standard argument  shows that the {existence} of a certain polynomial {of the form} 
\begin{equation}
Q(\theta)=Q(t,s) = \sum_{m,n=1}^M b_{m,n} \phi_{m}(t)\phi_n(s),
\label{eq:form of dual polynomial}
\end{equation}
would guarantee \editr{the success of Program~\eqref{eq:feas}  in the absence of noise.}
% that Program~\eqref{eq:feas}  successfully recovers a sparse nonnegative measure in the absence of noise. 
\edit{Known as the {dual certificate} for Program~\eqref{eq:feas}, this polynomial $Q$} has to be nonnegative on $\I^2$, with zeros only at the impulse locations $\Theta = \{\theta_k\}_{k=1}^K=\{(t_k,s_k)\}_{k=1}^K$. Setting $T=\{t_k\}_{k=1}^K$ and $S=\{s_k\}_{k=1}^K$ \editr{for short}, the proof \editr{then} constructs \editr{the polynomial} $Q$ by carefully combining nonnegative univariate polynomials \edit{with prescribed zeros} on subsets of $T$ and $S$. 
% For example, we will need to construct a nonnegative polynomial $q_T(t)=\sum_{m=1}^M b_m \phi_m(t)$ with zeros only at $T$, {and similarly for $q_S(s)$ with zeros only at $S$. 
 In turn, such univariate polynomials exist if $\{\phi_m\}_{m=1}^M$ form a $\mathcal{C}$-system on \editr{the interval} $\I$;  see Section \ref{sec:nonoise proof} for the details.  The basic idea of the proof is visualized in Figure~\ref{fig:idea of proof noise free}. 
 
\end{rem} 
%  \edita{It is also worth noting that successful noise-free recovery may also be guaranteed with somewhat simpler constructions for the dual certificate, as outlined after Lemma \ref{lem:dual cert required} below. However, it appears that these simpler dual constructions lead to suboptimal number of measurements, i.e., $M=O(K^2)$ as opposed to $M=O(K)$ above. This, in a sense, justifies the more intricate construction in the proof of Proposition~\ref{prop:noisefree}. }
 
\subsection{Arbitrary Measure With Noise \label{sec:With-Noise}}

In this section, we present {the main result of this paper}. 
Theorem~\ref{thm:main noisy simplified} below generalizes Proposition~\ref{prop:noisefree}  to account for \circled{1}~model mismatch, where $x$ is not necessarily a well-separated sparse measure but might be close to one,  and \circled{2}~imaging noise~($\delta \ge 0$). \editr{That is, Theorem~\ref{thm:main noisy simplified} below}  addresses the stability of Program \eqref{eq:feas} to model mismatch and its robustness against imaging noise. 
Some preparation is necessary before presenting the result.

\subsubsection{Separation }
Unlike sparse and noise-free super-resolution in Proposition \ref{prop:noisefree}, a notion of separation \editr{will play} a role in Theorem~\ref{thm:main noisy simplified}. 

\editr{
\begin{defn}[Separation]\label{defn:separation}
For an atomic measure $x$ supported on $\Theta=\{\theta_k\}_{k=1}^K=\{(t_k,s_k)\}_{k=1}^K \subset \I^2$, let $\text{sep}(x)$  \editr{denote} the minimum separation between all impulses in $x$  and the boundary of $\I^{2}$. That is, $\mathrm{sep}(x) $ is the largest number $\nu$ such that
\[
\nu\le\left|t_{k}-t_{l}\right|,\qquad\nu\le\left|s_{k}-s_{l}\right|,\qquad k\ne l,\,\,k,l\in[K],
\]
\[
\nu\le\left|t_{k}-0\right|,\qquad\nu\le\left|t_{k}-1\right|,
\]
\begin{equation}
\nu\le\left|s_{k}-0\right|,\qquad\nu\le\left|s_{k}-1\right|.\label{eq:min sep}
\end{equation}
Naturally, if the measure $x$ satisfies $\text{sep}(x) = \varepsilon$, we call $x$ an {$\varepsilon$-separated measure.} 

\end{defn}
}
For example,  for $x$ in Figure \ref{fig:impulseLocations}, we have  $\text{sep}(x)=\min(t_2-t_1,s_2-s_1,t_1,1-t_2,s_1,1-s_2)$. 
\edita{\editr{We remark that} \editr{our} notion of separation is more restrictive than the one} commonly used in the super-resolution literature~\cite{Candes2014,bendory2016robust,tang2013compressed},  \edita{as it requires the point sources to be separated in both $t$ and $s$ directions by at least $\nu$.}

\subsubsection{Generalized Wasserstein distance} 
As an error metric, we \editr{will} use the generalized Wasserstein distance~\cite{piccoli2012generalized}, \edita{which is closely related to the notion of unbalanced transport \cite{chizat2018scaling}.}  
{We first recall  the total-variation (TV) norm of a measure on $\I^2$~\cite{villani2008optimal}} is defined as {\color{black}$\|z\|_{\mathrm{TV}}= \int_{\I^2} |z(\der t)|$}, akin to $\ell_1$-norm in finite dimensions.
\editr{Recall also that} the Wasserstein distance~\editr{\cite{villani2008optimal}} for two nonnegative measures \editr{$z_1$ and $z_2$, supported} on $\I^2$, is defined as
\begin{equation}
d_{\W}\left(z_1,{z}_2\right)=\inf\int_{\I^2\times \I^2}\left\|\tau_1-{\tau_2}\right\|_1 \cdot \gamma\left(\der \tau_1,\der {\tau_2}\right),
\label{eq:def of EMD}
\end{equation}
where the infimum is over every nonnegative  measure $\gamma$
on $\I^2\times \I^2$ that produces $z_1$ and ${z}_2$ as marginals, i.e.,
\begin{equation}
z_1(A_1) = \int_{A_1\times \I^2} \gamma(\der  \tau_1,\der \tau_2),
\qquad 
z_2(A_2) = \int_{\I^2 \times A_2} \gamma(\der \tau_1,\der \tau_2),
\end{equation}
for all measurable sets $A_1,A_2\subseteq \I^2$. If we were to think of $z_1,z_2$ as two piles of dirt, then $d_{\W}(z_1,z_2)$ \editr{is} the least amount of work needed to transform \editr{one pile to the other}. 
The Wasserstein distance is defined only if the TV norms of the two measures are equal, \editr{i.e., $\| z_1\|_{\mathrm{TV}} = \|z_2\|_{\mathrm{TV}} $.}  The generalized Wasserstein distance \edit{extends} $d_\W$ \editr{and} allow\editr{s} for calculating the distance between nonnegative measures with different TV norms. 

\editr{
\begin{defn}[Generalized Wasserstein distance]\label{defn:genWDist} For two nonnegative measures $x_1$ and $x_2$ supported on $\I^2$, their generalized Wasserstein distance is defined \editr{as}
\begin{equation}
d_{\G}\left(x_1,x_2\right)=\inf \l(\l\| x_1-z_1\r\|_{\mathrm{TV}}+d_{\W}\left(z_1,{z_2}\right) +\l\| {x_2}-{z_2}\r\|_{\mathrm{TV}}\r),
\label{eq:def of gen EMD}
\end{equation}
where the infimum is over every pair of nonnegative Borel measures $z_1$ and ${z}_2$ supported on $\I^2$ such that $\|z_1\|_{\mathrm{TV}}=\|{z}_2\|_{\mathrm{TV}}$.
\end{defn}
}

\editr{Compared to~\eqref{eq:def of EMD}}, the two new terms \editr{in \eqref{eq:def of gen EMD}} gauge the  difference between \editr{the mass of} $x_1$ and \editr{the mass of} $x_2$. 
\edita{Our choice of error metric \editr{$d_{\G}$} is  \editr{natural} in the sense that \editr{any} solution of Program~\eqref{eq:feas} is itself a measure. However, \editr{note that} controlling \editr{the} error  in the space of measures \editr{in our main result below} does not immediately translate \editr{in}to controlling the error in the location of point sources, which may be estimated by, \editr{for example,} applying the Prony's method~\cite{weiss1963prony} to a solution of Program~\eqref{eq:feas}.  

For \editr{instance}, when $\varepsilon$ is \editr{infinitesimally small}, \editr{the measure} $\delta_{1/2}+\delta_{1/2+\varepsilon}$ has a small distance in the Wasserstein metric from \editr{the measure} $2\delta_{1/2}$, \editr{but} it is indeed impossible to distinguish the \editr{two impulses} in the presence of noise in general, unless we impose additional structure on the noise. Nevertheless, in the limit of vanishing noise, it is possible to directly control the error in the location of point sources and we refer the reader to~\cite{poon2017multi,denoyelle2015support,Duval2015} for \editr{the details}.}
% \tb{I think the entire last paragraph should be removed}

\subsubsection{Model mismatch}
 \editr{Our main result, Theorem~\ref{thm:main noisy simplified} below}, bounds the recovery error $d_{\G}(x,\widehat{x})$, where $\widehat{x}$ is a solution of Program~\eqref{eq:feas}. 
\editr{Note that,} even though $x$ is an arbitrary nonnegative measure in this section, it can always be approximated with a well-separated sparse measure, up to some error with respect to the metric $d_{\G}$. \editr{This sparse measure will play a key role in Theorem~\ref{thm:main noisy simplified}.} 

\editr{
\begin{defn}[Residual]For a nonnegative measure $x$ supported on $\I^2$, given an integer $K$ and  $\varepsilon\in (0,1/2]$, there exists a  $K$-sparse nonnegative measure $x_{K,\varepsilon}$ that is  $\varepsilon$-separated and \editr{well-}approximates $x$. More specifically, {for any $\varepsilon\in (0,1/2]$,} there exists a $K$-sparse and $\varepsilon$-separated  nonnegative measure $x_{K,\varepsilon}$ 
such that 
\begin{equation}
R(x,K,\varepsilon):= d_{\G}\l (x,x_{K,\varepsilon}\r) = \min d_{\G}(x,\chi),
\label{eq:residual}
\end{equation}
where the {minimum} above is over every nonnegative  $K$-sparse  and $\varepsilon$-separated measure $\chi$ supported on $\mbox{interior}(\I^2)$. 
\end{defn}
}

\editr{In words,} the residual $R(x,K,\varepsilon)$ {can be thought of as} the mismatch in modelling $x$ with a well-separated sparse measure. 
 Indeed, note that the minimum in~\eqref{eq:residual} is   achieved: We can limit the search in  \eqref{eq:residual} to the (bounded) set of $K$-sparse and $\varepsilon$-separated measures with TV norm bounded by $\|x\|_{\mathrm{TV}}$. This set is also closed, \editr{with respect to the weak topology imposed by $d_{\G}$, see~\cite[Theorem~13]{piccoli2012generalized}, }
 and thus compact. Lastly, the objective function $d_{\G}$ of \eqref{eq:residual} is a norm and thus continuous a fortiori, hence the claim. 
%  \tb{Comment: I think this entire paragraph is taken from your 1-D paper, right? I don't fully understand the topological notions in the reference so I tried to be as minimal as possible.}
%The case $\lambda=1$ is excluded here because the infimum on the far-right of~\eqref{eq:residual} might not be achieved. \edit{In what follows, }{the dependence} of $R(x,K,\epsilon)$ on $\lambda$ is suppressed to ease the notation. 

%\paragraph{Imaging apparatus.} 

\subsubsection{Smoothness}
For Program~\eqref{eq:feas} to succeed in the general settings of this section, \edit{we \editr{also} impose additional \editr{requirements} on the imaging apparatus in the next two paragraphs.}
We assume \editr{in this section} that \editr{the imaging apparatus is smooth in the following sense.}

\editr{
\begin{defn}[Smoothness]
The imaging apparatus in~\eqref{eq:meas model} is $L$-Lipschitz-continuous if 
\begin{equation}
\l\| \int_{\I^2} \Phi(\theta) (x_1(\der\theta)-x_2(\der\theta))  \r\|_{\mathrm{F}}
\le L \cdot d_{\G}(x_1,x_2),
\label{eq:Lipschitz assumption}
\end{equation}
for every pair of measures $x_1,x_2$ supported on $\I^2$.
\end{defn}
}

 It is \editr{often} not difficult to verify the  Lipschitz-continuity of $\Phi$ with respect to $d_{\G}$, \editr{as the following example demonstrates.}

\begin{example}[Smoothness]
\editr{As a toy} example, suppose for simplicity that $\|x_1\|_{\mathrm{TV}}=\|x_2\|_{\mathrm{TV}}$, so that $d_{\G}(x_1,x_2) = d_\W(x_1,x_2)$, see~\eqref{eq:def of gen EMD}. Moreover, for clarity, let $m=1$ and \editr{note that}
\begin{align}
\int \Phi(\theta) (x_1(\der \theta) - x_2(\der \theta) ) = \int \phi(t)\phi(s) (x_1(\der t,\der s) - x_2(\der t,\der s) ),
\label{eq:simple1}
\end{align}
where $\phi$ is the measurement window. Let also $L_\phi $ denote the Lipschitz constant of $\phi(t)\phi(s)$ with respect to  $\ell_1$-norm, i.e.,
\begin{align}
\l| \phi(t_1)\phi(s_1) - \phi(t_2)\phi(s_2) \r| \le L_\phi (|t_1 - t_2|+ |s_1-s_2|),
\end{align} 
for every $t_1,t_2,s_1,s_2\in \mathbb{I}$. Then, recalling the Kantorovich duality~\cite{villani2008optimal}, we may write that 
\begin{align}
\l\| \int_{\mathbb{I}^2} \Phi(\theta) (x_1(\der \theta )-x_2(\der \theta) ) \r\|_{\mathrm{F}} &
= \l| \int_{\mathbb{I}} \phi(t)\phi(s) (x_1(\der t,\der s)-x_2(\der t,\der s))\r| 
\qquad \editr{\text{(see \eqref{eq:simple1})}}
\nonumber\\
 & = L_\phi \l| \int_{\mathbb{I}} \frac{\phi(t)\phi(s)}{L_\phi} (x_1(\der t,\der s)-x_2(\der t,\der s))\r| \nonumber\\
& \le L_\phi \cdot \max_{\psi} \int \psi(\theta) x_1(\der \theta) - \psi(\theta) x_2(\der \theta) \nonumber\\
& = L_\phi \cdot d_\W(x_1,x_2) 
\qquad \text{(Kantorovich duality)}
\nonumber\\
& = L_\phi \cdot  d_{\G}(x_1,x_2),
\qquad \text{(simplifying assumption)}
\end{align}
where the maximum \editr{in the third line} above is over all $1$-Lipschitz-continuous functions $\psi:\mathbb{I}^2\rightarrow\mathbb{R}$ with respect to $\ell_1$-norm. We conclude that \eqref{eq:Lipschitz assumption} holds with $L=L_\phi$ in this example. \hfill$\blacksquare$
\end{example}

\subsubsection{$\mathcal{C}^*$-system}
\begin{figure}[ht!]
\begin{center}
\includegraphics[width=.7\textwidth]{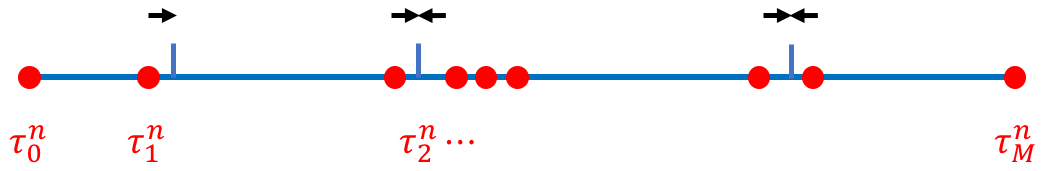}
\caption{This figure \editr{illustrates} an example of  an admissible sequence; see Definition~\ref{def:admissible seq}. For \editr{a} {fixed} $n$, the red dots form the \editr{increasing} sequence $\{\tau_k^n\}^{M}_{k=0}$. Note that the end points of the interval are included in the sequence\edit{, i.e., $\tau_0^n = 0$ and $\tau^n_M=1$.} \editr{In view of  Definition~\ref{def:admissible seq},} as $n\rightarrow\infty$, the  sequence $\{\tau_k^n\}_{k=1}^{M-1}$ converges element-wise to three distinct points on the interior of the interval, shown with blue bars. \editr{Moreover,}  all limit points \edit{in $(0,1)$} have an even multiplicity except for one, which has a multiplicity of exactly one.   \label{fig:admissible}} 
\end{center}
\end{figure}

To study the stability of Program \eqref{eq:feas}, we also need to modify the notion of $\mathcal{C}$-system in Definition~\ref{def:(T-systems)-Real-valued-and}. \edit{We begin} with the definition \editr{of an admissible sequence, visualized in Figure~\ref{fig:admissible}.}
\begin{defn}[Admissible sequence]\label{def:admissible seq} 
\emph{
\editr{For a pair of integers $K$ and $M$ obeying $M\ge 2K+1$,} we say that $\{\{\tau_{k}^{n}\}_{k=0}^M\}_{n\ge 1}\subset \I$ is a  $(K,\varepsilon)$-{admissible} sequence if:
\begin{enumerate}[leftmargin=*]
\item
$\tau_{0}^{n}=0$ and $\tau_{M}^{n} = 1$  for every  $n$, i.e., the endpoints of $\I=[0,1]$ are included in \edit{the} \editr{increasing} sequence $\{\tau_k^n\}_{k=0}^M$, \edit{for every $n$}. 
\item As $n\rightarrow\infty$, \edit{the \editr{increasing} sequence} $\{\tau_{k}^{n}\}^{M-1}_{k=1}$ converges \edit{(element-wise)} to  an $\varepsilon$-separated \editr{finite} subset of $\I$ with at most $K$ \edita{distinct} points, where every element has an even multiplicity, except one element that appears only once.\footnote{{That is, every element is repeated an even number of times ($2,4,\cdots$) except one element that appears only once.}}
\end{enumerate}
}
\end{defn}
% An example of admissible sequences is given in Figure \ref{fig:admissible} \tb{You already said that before the Definition}. 
\edit{While $\mathcal{C}$-system in Definition~\ref{def:(T-systems)-Real-valued-and} is a condition on all increasing sequences of length $M$, the $\mathcal{C}^*$-system below is a condition only on admissible sequences; these are the only sequences that matter in our analysis.}  
Like a $\mathcal{C}$-system, \edit{a} $\mathcal{C}^*$-system imposes certain requirements on a family of functions. Whereas the performance of Program \eqref{eq:feas} for sparse measures and in the absence of noise relates to a certain $\mathcal{C}$-system in \editr{Proposition~\ref{prop:noisefree}}, the general performance of Program~\eqref{eq:feas} relates to certain $\mathcal{C}^*$-systems, as we will see shortly in Theorem~\ref{thm:main noisy simplified}. \edit{The definition of $\mathcal{C}^*$-system below is immediately followed by its motivation.} 
\begin{defn}[$\mathcal{C}^*$-system]
\label{def:(T-systems,-modified)-For}
\emph{
\editr{For an integer $K$ and an even integer $M$ \editr{obeying} $M\ge 2K+2$,} real-valued functions $\{\phi_m\}_{m=0}^M$ form a $\mathcal{C}^*_{K,\varepsilon}$-system on $\I$ if \editr{every} $(K,\varepsilon)$-admissible sequence $\{\{\tau_{k}^n\}_{k=0}^{M}\}_{n\ge 1}$ satisfies:} 
\emph{
\begin{enumerate}[leftmargin=*]
\item The determinant of \edita{the $(M+1)\times (M+1)$ matrix  $[\phi_{m}(\tau_{k}^n)]_{k,m=0}^M $} is  \editr{positive for all sufficiently large $n$.} 
% \editr{asymptotically. That is, {$   \det ( [\phi_{m}(\tau_{k}^n)]_{k,m=0}^M ) \ge 0$ for all sufficiently large $n$.} \tb{remove all}} 
\item  Moreover, all minors along the $\underline{l}$th row of \editr{the matrix} $[\phi_{m}(\tau_{k}^n)]_{k,m=0}^M$ approach zero at the same rate when $n\rightarrow\infty$. 
Here, $\underline{l}$ is the index of the element of  the limit sequence that appears only once.\footnote{A nonnegative sequence $\{u^n\}_{n\ge 1}$ approaches zero at the rate $n^{-p}$ if $u_n=\Theta(n^{-p})$.  See, for example, \editr{page 44 of }\cite{cormen2009introduction}.}
\end{enumerate}
}
\end{defn}

 \editr{
 \begin{rem}[Properties of $\mathcal{C}^*$-systems]\label{rem:propsTStarSys} 
 \circled{1}~Note that a $\mathcal{C}^*_{K,\varepsilon}$-system on $\I$ is also a $\mathcal{C}^*_{K',\varepsilon}$-system  for every integer $K'\le K$. Indeed, this claim follows from the observation that every $(K',\varepsilon)$-admissible sequence is itself a $(K,\varepsilon)$-admissible sequence.
 \circled{2}~Moreover, if $\{\phi_m\}_{m=0}^M$ form a $\mathcal{C}^*_{K,\varepsilon}$-system on $\I$, then so do the scaled functions $\{c_m \phi_m\}_{m=0}^M$ for positive constants $\{c_m\}_{m=0}^M$.
 \end{rem}
 }

\edit{Let us \editr{also offer}  some \editr{insight about} $\mathcal{C}^*$-systems. 
In the proof of Proposition \ref{prop:noisefree} for  sparse and noise-free super-resolution, in order to \editr{construct} a polynomial 
$$
\sum_{m=1}^M b_m \phi_m \ge 0,
$$
with prescribed zeros on $\I$, we \editr{require} that $\{\phi_m\}_{m=1}^M$ form a  $\mathcal{C}$-system; see the discussion after Definition~\ref{def:(T-systems)-Real-valued-and}. On the other hand, \editr{for a given function $\phi_0$ (which, in our context, will signify the stability to noise in super-resolution), in order} to \editr{construct} a polynomial 
$$
\sum_{m=1}^M b_m \phi_m  \ge \phi_0,
$$
\editr{where equality holds at} prescribed \editr{points in}  $\I$, \editr{it is natural to require} $\editr{\{\phi_0\}} \cup \{\phi_m\}_{m=1}^M$ to form a $\mathcal{C}$-system. The definition of $\mathcal{C}^*$-system above is based on the same idea but limited to admissible sequences, to ease the burden of verifying the conditions in Definition~\ref{def:(T-systems,-modified)-For}. In particular, Definition~\ref{def:(T-systems,-modified)-For} \editr{will help} exclude  trivial polynomials, such as $0\cdot \phi_0 + \sum_{m=1}^M b_m \phi_m$. 
% \editr{and ensures that the coefficient of $\phi_0$ is negative. \tb{the last addition is unclear}} 
}

% Let us \editr{further} compare T- and T$^*$-systems. 
% An arbitrary polynomial of a  T-system has a limited number of zeros, see the discussion after Definition \ref{def:(T-systems)-Real-valued-and}.
%  Polynomials of a T$^*$-system, on the other hand, have no such property since the determinant in  part 1 of Definition \ref{def:(T-systems,-modified)-For} might vanish on $\I$.

% Instead, the notion of T$^*$-system is designed to facilitate the construction of the necessary dual certificates for Program~\eqref{eq:feas} in the presence of model mismatch and noise.
% In particular, \edit{as mentioned earlier,} Part~2 in Definition~\ref{def:(T-systems,-modified)-For} is designed to exclude trivial polynomials that  do not qualify as dual certificates.   
\editr{We remark that} Definition~\ref{def:(T-systems,-modified)-For} only considers admissible sequences to simplify the burden of verifying whether a family of functions form a $\mathcal{C}^*$-system.

To summarize, the widely-used notion of $\mathcal{C}$-system in Definition \ref{def:(T-systems)-Real-valued-and} plays a key role in the analysis of sparse inverse problems in the absence of noise, whereas $\mathcal{C}^*$-system above \editr{was} introduced in \cite{eftekhari2018sparse} and tailored for the {stability analysis} of sparse inverse problems.\footnote{\editr{Let us point out that we use the shorthand of $\mathcal{C}^*$-system here instead of  T$^*$-system used in~\cite{eftekhari2018sparse}.}} It was established in~\cite{eftekhari2018sparse} that translated copies of the Gaussian window $e^{-t^2}$ indeed form a $\mathcal{C}^*$-system, under mild conditions \editr{reviewed} in Section~\ref{sec:gaussian} below. We \editr{suspect} this to also hold for many other measurement windows with sufficiently fast decay.\footnote{The definition of \editr{$\mathcal{C}^*$-system} here is slightly different from that in \cite{eftekhari2018sparse} but the difference is inconsequential. 
% The new definition here will help improve the dependence of recovery error bounds on number of observations both in one- and two-dimensional problems.
} 

\subsubsection{Main Result}
We are now ready to present \editr{the} main result \editr{of this paper,} \editr{which quantifies} the performance of Program~\eqref{eq:feas} in the general case where $x$ is an arbitrary nonnegative measure on $\I^2$ and in the presence of additive noise. Theorem~\ref{thm:main noisy simplified}, proved in Section \ref{sec:noisy}, states that Program~\eqref{eq:feas} approximately recovers $x$ provided that certain $\mathcal{C}$- and $\mathcal{C}^*$-systems exist. \edit{As an example of this  general result, Section \ref{sec:gaussian} later \editr{specializes} Theorem \ref{thm:main noisy simplified} \editr{for} imaging \editr{under} Gaussian blur.}

\begin{thm}[Arbitrary measure with noise]
\label{thm:main noisy simplified} Consider a nonnegative measure $x$ supported on $\I^2$. Consider also a noise level $\delta\ge 0$, measurement functions $\{\phi_m\}_{m=1}^M$, 
% the corresponding imaging operator $\Phi$, 
and the image $y\in\mathbb{R}^{M\times M}$, \editr{see~\emph{(\ref{eq:shorthands},\ref{eq:meas model})}}. \editr{We assume that the imaging apparatus is $L$-Lipschitz in the sense of~\eqref{eq:Lipschitz assumption}.}   

For an integer $K$ and $\varepsilon\in (0,1/2]$, 
let $x_{K,\varepsilon}$ be a  $K$-sparse and $\varepsilon$-separated  nonnegative measure on $\I^2$ that approximates $x$ \editr{with the residual of $R(x,K,\varepsilon)$}, in the sense of~\eqref{eq:residual}. In particular, let $\Theta=\{\theta_k\}_{k=1}^K =\{(t_k,s_k)\}_{k=1}^K\subset \interior(\I^2)$ \editr{denote} the support of $x_{K,\varepsilon}$, and set $T=\{t_k\}_{k=1}^K$ and $S=\{s_k\}_{k=1}^K$ for short.  

\edit{With~$\widehat{x}$ denoting a solution of Program~\eqref{eq:feas} for $\delta'\ge (1+L\cdot  R(x,k,\varepsilon))\delta$, it holds that 
\begin{align}
 d_{\G}\left( x,\widehat{x}\right) 
& \le c_1  \delta+ \editr{c_2(\varepsilon)} + c_3  R(x,K,\varepsilon),
\label{eq:EMD simplified}
\end{align}
where $d_{\G}$ is the generalized Wasserstein metric in~\eqref{eq:def of gen EMD}.  \editr{Above, $c_1,c_2(\varepsilon),c_3$} are specified explicitly in~\eqref{eq:c3}, and depend on the the measure $x$, the separation $\varepsilon$, and  the \editr{measurement functions $\{\phi_m\}_{m=1}^M$}. \editr{In particular, it holds that $c_2(0)=0$.}

The error bound in \eqref{eq:EMD simplified}  \editr{holds} if $M\ge2K+2$ \editr{and}
\begin{enumerate}[leftmargin=*]
\item  $\{\phi_{m}\}_{m=1}^M$ form a $\mathcal{C}$-system on $\I$,
\item $\{F_{T_\Omega}\}\cup\{\phi_{m}\}_{m=1}^M$ and $\{F_{S_\Omega}\}\cup\{\phi_{m}\}_{m=1}^M$ both form  $\mathcal{C}^*_{K,\varepsilon}$-systems on $\I$ for every $\Omega\subseteq [K]$, 
\item  \editr{$\{F_{t_k}^+\}\cup\{\phi_{m}\}_{m=1}^M$ and $\{F_{t_k}^-\}\cup\{\phi_{m}\}_{m=1}^M$} both  form  $\mathcal{C}^*_{K,\varepsilon}$-systems on $\I$ for every $k\in [K]$,
\item  \editr{$\{F_{s_k}^+\}\cup\{\phi_{m}\}_{m=1}^M$} form  a $\mathcal{C}^*_{K,\varepsilon}$-system on $\I$ for every $k\in [K]$. 
\end{enumerate} 
\editr{Above,} for every index set $\Omega\subseteq [K]$ and $k\in [K]$, we define the functions \editr{$F_{T_\Omega},F_{S_\Omega},F_{t_k}^{\pm},F_{s_k}^+: \I \rightarrow \mathbb{R}$} as
\[
F_{T_\Omega}(t):=\begin{cases}
0, & \text{when there exists }k\in \Omega \text{ such that } |t-t_k| \le\varepsilon/2,\\
\pline, & \text{elsewhere on }\interior(\I),
\end{cases}
\]
\[
F_{S_\Omega}(s):=\begin{cases}
0, & \text{when there exists }k\in\Omega \text{ such that }|s-s_k| \le {\varepsilon}/2,\\
\pline, & \text{elsewhere on }\interior(\I),
\end{cases}
\]
\editr{
\[
 F_{t_k}^{\pm}(t):=\begin{cases}
 \pm 1, & \text{when } |t-t_k| \le \varepsilon/2 ,\\
0, & \text{everywhere else on }\interior(\I),
\end{cases}
\]
\[
 F_{s_k}^+(s):=\begin{cases}
1, & \text{when } |s-s_{k}|\le \varepsilon/2,\\
0, & \text{everywhere else on }\interior(\I).
\end{cases}
\]
}
}
\end{thm}

\begin{figure}[H]
\begin{center}

\subfloat[\label{fig:lowerBnd}]{\protect\includegraphics[width=0.28\textwidth]{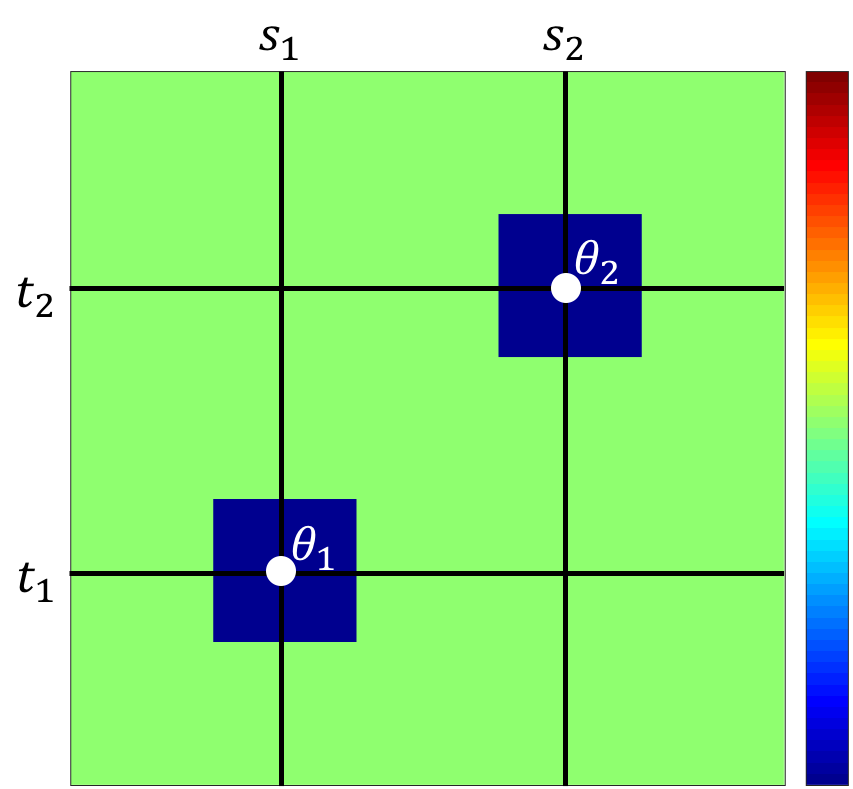}
}
\subfloat[\label{fig:noisyPoly1}]{\protect\includegraphics[width=0.28\textwidth]{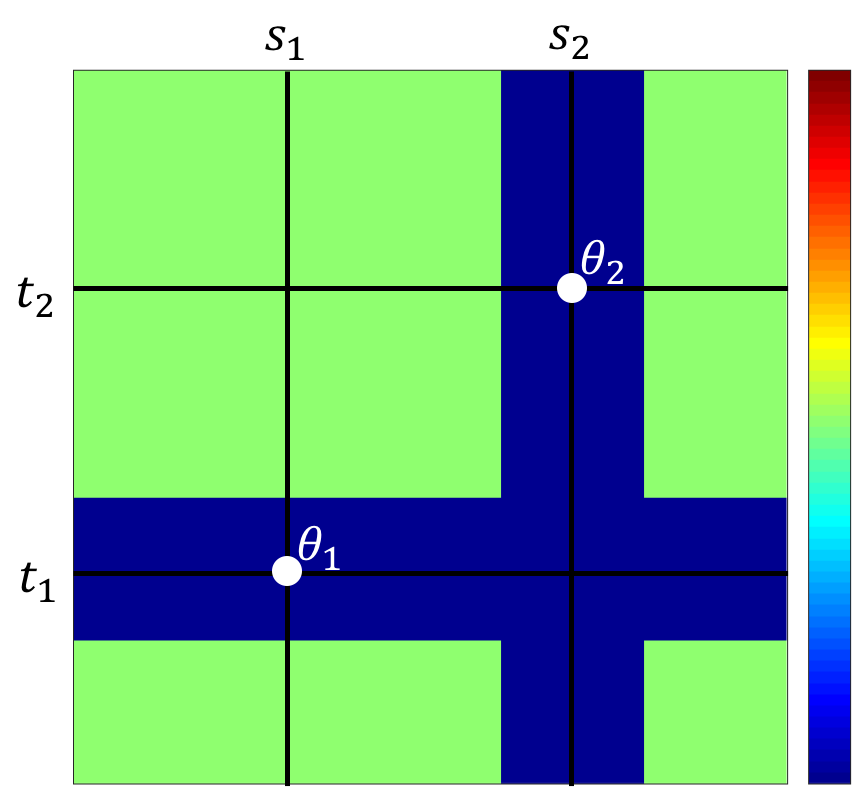}
}
\subfloat[\label{fig:noisyPoly2}]{\protect\includegraphics[width=0.28\textwidth]{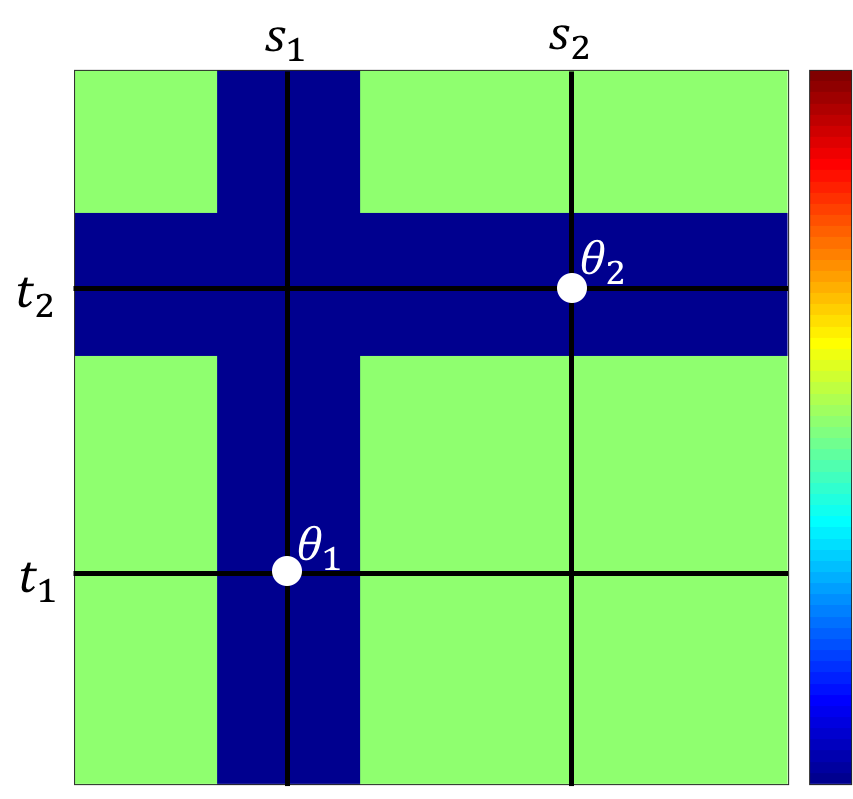}
}

\subfloat[\label{fig:noisyPolySum}]{\protect\includegraphics[width=0.28\textwidth]{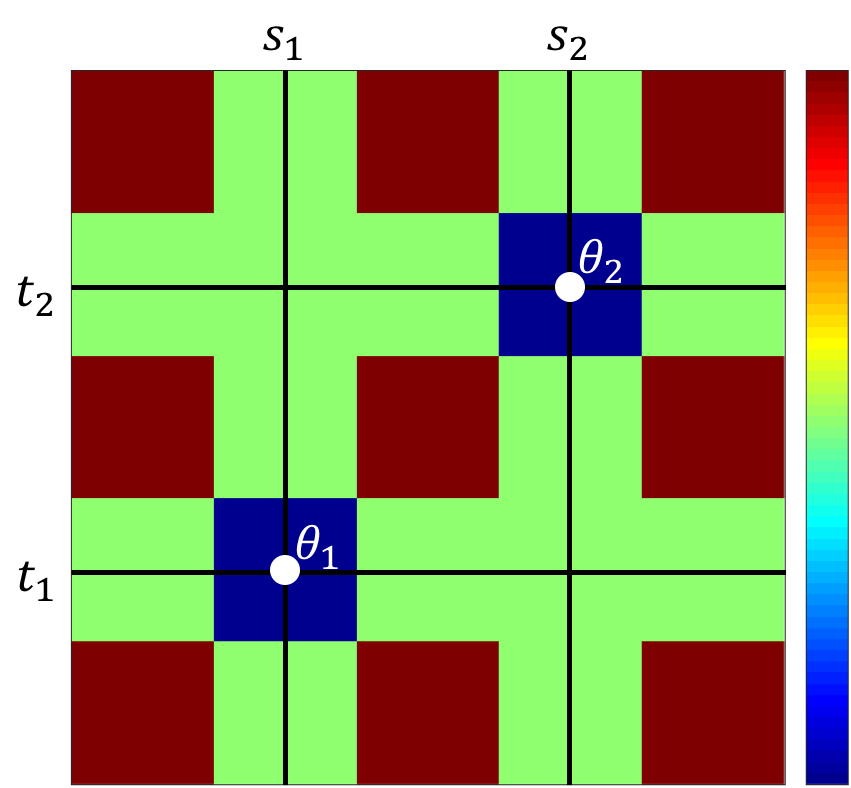}
}
\subfloat[\label{fig:noisyDual1d}]{\protect\includegraphics[width=0.28\textwidth, height=.2363\textwidth]{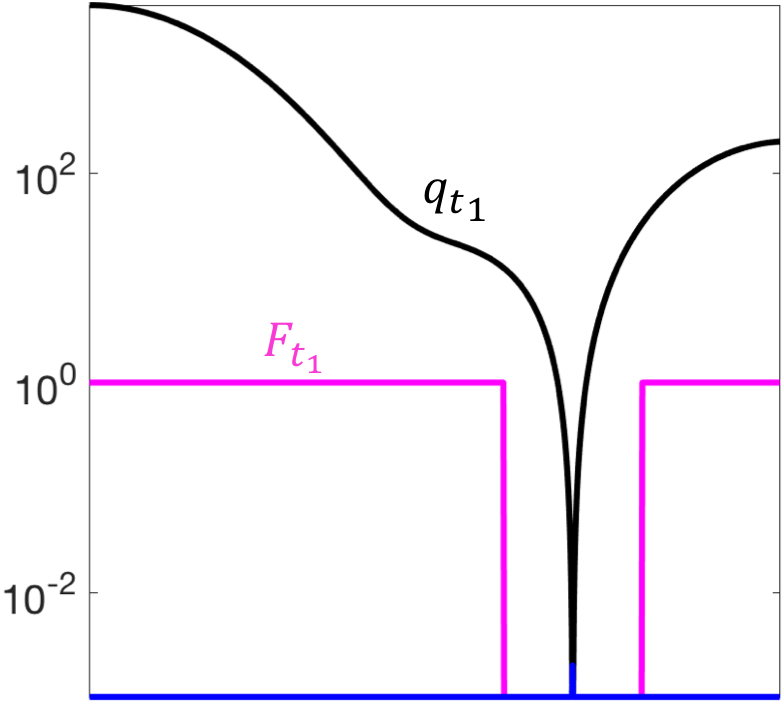}
}
\subfloat[\label{fig:noisyDual2d}]{\protect\includegraphics[width=0.28\textwidth]{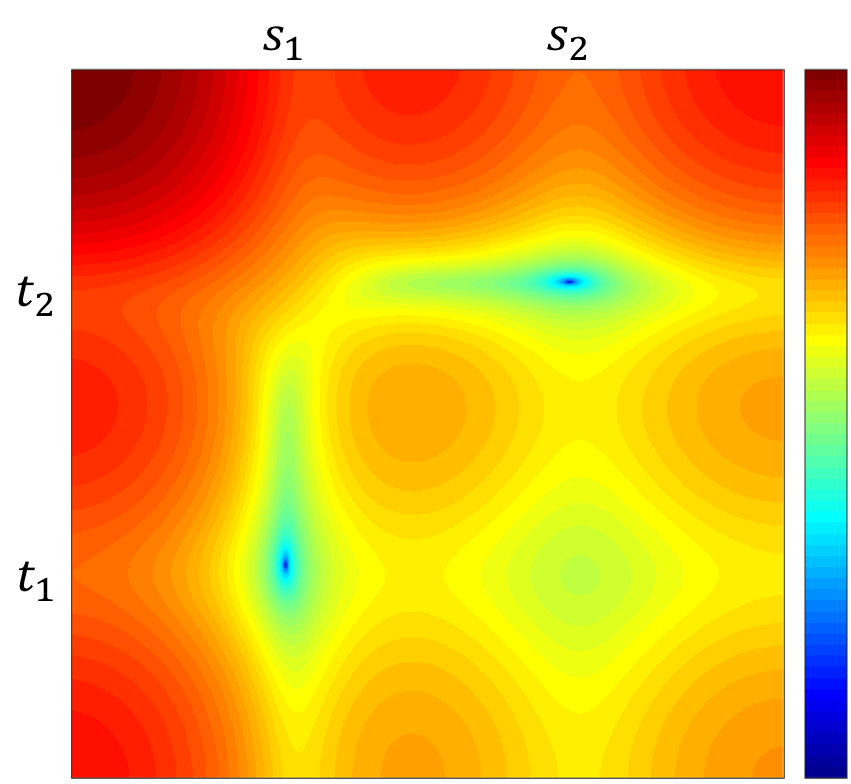}
}
\caption{\small{This figure \editr{visualizes} some of the \editr{principle} ideas behind the proof of Theorem \ref{thm:main noisy simplified}, \editr{explained in Remark~\ref{rem:proofTech2}.} As an example, consider the nonnegative measure $x=a_1\delta_{\theta_1}+a_2 \delta_{\theta_2}$. The locations of two impulses at $\theta_1=(t_1,s_1)\in \I^2$ and $\theta_2 = (t_2,s_2)\in \I^2$ are shown with white dots, and the black lines show the corresponding grid. 
For Program \eqref{eq:feas} to approximately recover $x$ from the noisy image $y$ \edit{given} in \eqref{eq:noise}, we need to construct a nonnegative polynomial $Q$ of the form in \eqref{eq:form of dual polynomial} that is zero at the impulse locations and large away from the impulses, to ensure stability. That is, we need $Q(\theta_1)=Q(\theta_2)=0$,  and also $Q(\theta)\ge \overline{g}>0$ when $\theta$ is far from both $\theta_1,\theta_2$, for a positive \editr{scalar} $\overline{g}$.  This lower bound for \editr{the} polynomial $Q$ is shown in Figure~\ref{fig:lowerBnd} as a heat map, with warmer colors corresponding to larger values, i.e., blue corresponds to \edit{zero} and green corresponds to $\overline{g}$. \\ Let us first express this lower bound on $Q$ in terms of univariate functions. To that end, consider a function $F_{t_1}(t)$ that is zero near and equal to $\sqrt{\overline{g}}$ away from $t_1$.
%=\sum_{m=1}^M b_m^{11} \phi_m(t)
 Likewise, consider similar nonnegative functions $F_{t_2}(t),F_{s_1}(s),F_{s_2}(s)$ that are zero near and equal to $\sqrt{\overline{g}}$ away from $t_2,s_1,s_2$, respectively.  Figure~\ref{fig:noisyPoly1} shows the heat map of  $F_{t_1}(t)F_{s_2}(s)$, Figure~\ref{fig:noisyPoly2} shows the heat map of $F_{t_2}(t)F_{s_1}(s)$, and lastly Figure~\ref{fig:noisyPolySum} shows the heat map of their sum, i.e., $G(\theta) := F_{t_1}(t)F_{s_2}(s)+ F_{t_2}(t)F_{s_1}(s)$. Note that  $G$ is zero near and larger than $\overline{g}$ away from the impulse locations $\theta_1,\theta_2$, as desired. 
\\ It only remains to construct \editr{the} univariate polynomials $q_{t_1},q_{t_2},q_{s_1},q_{s_2}\in \text{span}(\phi_1,\cdots,\phi_M)$ that satisfy the inequalities $q_{t_1}\ge F_{t_1}$, $q_{t_2}\ge F_{t_2}$, $q_{s_1}\ge F_{s_1}$, $q_{s_1}\ge F_{s_1}$, \editr{with} equality at $t_1,t_2,s_1,s_2$, respectively. Under the \editr{assumptions of} Theorem~\ref{thm:main noisy simplified}, the existence of these univariate polynomials follows from~\cite{eftekhari2018sparse}. \editr{In this fashion}, we \editr{finally} obtain a nonnegative polynomial $Q(\theta) = q_{t_1}(t)q_{s_2}(s)+q_{t_2}(t)q_{s_1}(s)$ that is zero at the impulse locations and larger than $\overline{g}$ away from the impulses, as desired. For example, for the Gaussian window detailed in Section \ref{sec:gaussian} with \editr{the} {standard deviation} $\sigma=0.2$, Figure~\ref{fig:noisyDual1d} shows $q_{t_1}(t)$ and Figure~\ref{fig:noisyDual2d} shows the heat map of the dual certificate $Q(\theta)$, both in logarithmic scale. 
Yet, another polynomial $Q^0$ is needed to \editr{control the recovery error near the impulses and thus} complete the proof of Theorem~\ref{thm:main noisy simplified}, see Section~\ref{sec:noisy}. 
  \label{fig:idea of proof noisy}}} 
\end{center}
\end{figure}

\edit{
\editr{An example of the functions in Theorem~\ref{thm:main noisy simplified} appears in Figure~\ref{fig:noisyDual1d}, where the purple graph is an example of $F_{T_\Omega}$ for $\Omega = \{t_1\}$, shown in the figure  as $F_{t_1}$ for brevity.}
Theorem \ref{thm:main noisy simplified} for image super-resolution is unique in a number ways. The differences with prior work are further discussed  in Section \ref{sec:related} and also  summarized here. First, Theorem \ref{thm:main noisy simplified} applies to arbitrary measures, not only atomic ones. In particular, for atomic measures, no minimum separation or  limit on the density of impulses are imposed in contrast to earlier results  \cite{schiebinger2015superresolution,bendory2017robust,morgenshtern2014stable,denoyelle2015support}.

Moreover, Theorem \ref{thm:main noisy simplified} addresses both noise and model-mismatch in image super-resolution. Indeed, even in the 1-D case, stability was identified as a  technical obstacle in earlier work \cite{schiebinger2015superresolution}. In addition, the recovery error  in Theorem~\ref{thm:main noisy simplified} is quantified with a natural metric between measures, i.e., the generalized Wasserstein metric, in contrast to prior work; see for example~\cite{Fernandez-Granda2013} that separately studies the error near and away from the impulses. Lastly, the measurement functions $\{\phi_m\}_m$ are required to be continuous rather than (several times) differentiable \cite{schiebinger2015superresolution,denoyelle2015support}. All this is achieved without the need to explicitly regularize for sparsity in Program~\eqref{eq:feas}.

\edita{Note also that, in practice, we often have an upper bound for the noise level $\delta$ and the model mismatch $R(x,K,\epsilon)$, which would allow us to apply Theorem~\ref{thm:main noisy simplified}. This approach to quantifying stability against noise and model mismatch is common in model-based signal processing~\cite{candes2008introduction}.} 
% \tb{I think this paragraph can be removed}

Several \editr{additional} remarks are in order \editr{about}  Theorem~\ref{thm:main noisy simplified}. }

\begin{rem}[Proof technique]\label{rem:proofTech2} For Program~\eqref{eq:feas} to successfully recover a sparse measure in the absence of noise, {we} \editr{constructed} a nonnegative polynomial $Q(\theta)$, \edit{\editr{with}in the span of \editr{the} measurement functions}, \editr{which vanished} only at the impulse locations  $\Theta=\{\theta_k\}_{k=1}^K=\{(t_k,s_k)\}_{k=1}^K$, see the discussion after Proposition~\ref{prop:noisefree}.  For approximate recovery in the presence of model mismatch and noise,  we need to construct a nonnegative polynomial $Q(\theta)$ that is bounded \emph{away} from zero far from the impulse locations $\Theta$, i.e.,  
$$
Q(\theta) \ge \overline{g}>0,
\qquad \editr{\text{for every }} \theta \text{ far from } \Theta,
$$
\editr{where $\overline{g}$ is a positive scalar}. Letting $T=\{t_k\}_{k=1}^K$ and $S=\{s_k\}_{k=1}^K$ for \edit{short}, the proof of Theorem \ref{thm:main noisy simplified} constructs $Q$ by combining certain univariate polynomials, \edit{similar to} {the} proof of Proposition \ref{prop:noisefree} which was \editr{itself} summarized earlier in Section \ref{sec:noisefree}\editr{, and illustrated in}  Figure~\ref{fig:idea of proof noise free}. \edit{Among} these univariate polynomials, \edit{for example,} the proof constructs a nonnegative polynomial $q_T$ such that 
$$
q_T(t)\ge 1,
\qquad \editr{\text{for every }} t \text{ far from }T.
$$
\editr{As shown in~\cite{eftekhari2018sparse}, such a univariate polynomial $q_T$ exists if} $\{F_T\}\cup \{\phi_m\}_{m=1}^M$  form a $\mathcal{C}^*$-system.  In addition to $Q$, \edit{we also find it} necessary to construct yet another \editr{nonnegative} polynomial $Q^0$ \editr{to control the recovery error near the impulse locations and thus complete the proof of  Theorem~\ref{thm:main noisy simplified},} see Section~\ref{sec:noisy} for more details. Figure~\ref{fig:idea of proof noisy} illustrates some of the \editr{key} ideas in the proof of Theorem~\ref{thm:main noisy simplified}. 
\end{rem}

\begin{rem}[Recovery error] The \edit{bound on the} recovery error $d_{\G}(x,\widehat{x})$ in~\eqref{eq:EMD simplified} depends on the noise level $\delta$ and on how well $x$ can be approximated with a well-separated sparse measure. More specifically, for any $\varepsilon\in(0,1/2]$,   $x$ can  be approximated with a $K$-sparse and $\varepsilon$-separated measure $x_{K,\varepsilon}$, with \editr{the} residual of $R(x,K,\varepsilon)$, see \eqref{eq:residual}. We might then think of a solution $\widehat{x}$ of Program~\eqref{eq:feas} as an estimate for $x_{K,\varepsilon}$ and therefore an estimate for $x$, up to the residual $R(x,K,\varepsilon)$.  
Both the separation $\varepsilon$ and  the residual $R(x,K,\varepsilon)$ appear on the right-hand side of the error bound \eqref{eq:EMD simplified}.

In particular, when \editr{$\delta =\varepsilon= R(x,K,\varepsilon)= 0$}, we again obtain Proposition~\ref{prop:noisefree} for recovery of  \editr{a} $K$-sparse nonnegative measure in the absence of noise.
% provided that the coefficients $c_1,c_2,c_3$ remain bounded in the limit, which is indeed the case for the example of Gaussian windows  discussed in Section~\ref{sec:gaussian}. 
\edit{Note that, given a noise level $\delta$,} this work does not address \edit{the \editr{challenging} problem of  choosing} the separation $\varepsilon$ \edit{in order to minimize the right-hand side of \eqref{eq:EMD simplified}}. 
Intuitively, for \editr{large} $\delta$, we must choose \edit{the separation} $\varepsilon$ \editr{large} \editr{enough} \editr{to maintain stability against the large noise level}. In turn, \editr{a large} $\varepsilon$ leads to a \editr{large} residual $R(x,K,\varepsilon)$, \editr{see~\eqref{eq:residual}.} 
The correct balance between $\epsilon$ and $R(x,K,\epsilon)$ depends on the particular choice of the measurement functions $\{\phi_m\}_{m}$ and \editr{is beyond the scope of this paper}. 
\end{rem}

\begin{rem}[Minimum separation]
Theorem \ref{thm:main noisy simplified} applies to \emph{any} nonnegative measure $x$. In particular, when $x$ is an atomic measure, Theorem \ref{thm:main noisy simplified} applies \emph{regardless} of the separation between the impulses present in \editr{the measure} $x$. However, \editr{it is crucial to note that} the recovery error $d_{\G}(x,\widehat{x})$ does indeed {depend} on the separation of $x$. %, i.e., $\text{sep}(\Theta)$, where $\Theta$ is the support of $x$.

 As an example, consider \editr{the atomic measure}  $x=\delta_{0.5}+\delta_{0.51}$. In order to apply Theorem \ref{thm:main noisy simplified}, we can set $\varepsilon=0.01$, so that $x_{K,\varepsilon}=x$ and $R(x,K,\varepsilon)=0$. Now, \edit{the error bound in}~\eqref{eq:EMD simplified}  reads as 
\begin{equation}
d_{\G}(x,\widehat{x})  \le  \editr{c_1(0.01) \cdot \delta + c_2(0.01) },
\label{eq:case1}
\end{equation}
where \editr{we have made explicit the dependence of $c_1$ on $\varepsilon$ for emphasis.}
Alternatively, we may also apply Theorem \ref{thm:main noisy simplified} by setting $\varepsilon=0.2$, so that $x_{K,\varepsilon}=\delta_{0.405}+\delta_{0.605}$ and $R(x,K,\varepsilon) = 0.19$. In this case, \eqref{eq:EMD simplified} reads as 
\begin{equation}
d_{\G}(x,\widehat{x})  \le\editr{ c_1(0.2)\cdot  \delta + c_2(0.2)  + c_3(\varepsilon) \cdot 0.19.}
\label{eq:case2}
\end{equation}
% where $C'_1=c_1(x,\varepsilon)$, $C'_2=c_2(x,\varepsilon)$, and $C'_3=c_3(x,\varepsilon)$ with $\varepsilon=0.2$.  
% Informally speaking, \editr{in spite of the new third term on the right-hand side of~\eqref{eq:case2}}, one would expect the bound on recovery error in~\eqref{eq:case2} to be  smaller \editr{(better)} than the \editr{recovery error bound} in~\eqref{eq:case1}, because resolving nearby impulses is more difficult. 
% \tb{Not sure how did you get to this conclusion. Perhaps you refer only to the first two term?}
% and one would expect $c_2(x,\varepsilon)$ to implode as $\varepsilon\rightarrow 0$.  
% \edit{However, as mentioned in the previous paragraph, this} 
This
work, however, does not address the optimal choice of separation $\varepsilon$ as a function of noise level $\delta$. That is, given $\delta$, the choice of $\varepsilon$  that would minimize the right-hand of~\eqref{eq:EMD simplified} is not studied here\edit{; see also~\cite{eftekhari2018sparse}.} 
% \tb{remove the last sentence, you already said it}
\end{rem}

% \paragraph{Invariance.}
% Although not mentioned in Theorem \ref{thm:main noisy simplified}, as a sanity check, \edit{one might verify that the error bound on the} right-hand side of \eqref{eq:EMD simplified} is invariant under scaling of the noise level $\delta$ and the imaging operator $\Phi$. If we replace $\delta$ with $\alpha\delta$ and $\Phi$ with $\alpha \Phi$ for a positive $\alpha$, the right-hand side of \eqref{eq:EMD simplified} does not change; see~\cite{eftekhari2018sparse} for more details. 

\subsection{Example with Gaussian Window} \label{sec:gaussian}

As an  example of the general super-resolution framework presented in this paper, consider the case where $x$ is a $K$-sparse nonnegative measure as in \eqref{eq:atomic} and $\{\phi_m\}_{m=1}^M$ are translated copies of a one-dimensional Gaussian window, i.e.,
\begin{equation}
\phi_m(t) = g_1(t-t'_m) := e^{-(t-t'_m)^2/\sigma^2},
\label{eq:gauss window}
\end{equation}
for $T'=\{t'_m\}_{m=1}^M\subset \I$ and  {standard deviation} $\sigma>0$. \editr{Recalling~\eqref{eq:noise},} note that
\begin{align}
 \int_{\I^2} \phi_m(t) \phi_n(s) x(\der t,\der s) 
& = \sum_{k=1}^K a_k \cdot g_1(t_k-t'_m) g_1(t_k-t'_n)
\qquad \text{(see \eqref{eq:gauss window})}
%\qquad \l(\mbox{(\ref{eq:atomic},\ref{eq:gauss window}) and } \theta_k = (t_k,s_k) \r)
 \nonumber\\
& = \sum_{k=1}^K a_k \cdot g_2(\theta_k-\theta'_{m,n}) \nonumber\\
& = \sum_{k=1}^K a_k \cdot e^{-\frac{\|\theta_k-\theta'_{m,n}\|_2^2}{\sigma^2}},
\end{align}
{where $\theta_k = (t_k,s_k),  \theta'_{m,n} = (t'_m,t'_n)$ and $g_2$ is a 2-D Gaussian window, which can be thought of as the {point-spread function}} \editr{of \editr{the} imaging} \editr{device}.  Note that we might also think of $\{\theta'_{m,n}\}_{m,n=1}^M=T'\times T'$ as the sampling points in the sense that 
\begin{align}
\int_{\I^2} \phi_m(t) \phi_n(s) x(\der t,\der s)  = 
\sum_{k=1}^K a_k \cdot g_2(\theta_k-\theta'_{m,n}) = 
(g_2 \star x)(\theta'_{m,n}),
\qquad m,n\in [M],
\label{eq:sampling perspective}
\end{align}
{where $\star$ stands for convolution.
Put differently, the integral above evaluates the \editr{Gaussian-blurred (or filtered)} copy of measure $x$ at locations $T'\times T'$. 

Suppose first that there is no imaging noise  and, consequently, $y_{m,n}=(g_2\star x)(\theta'_{m,n})$ is the $(m,n)^{\text{th}}$ pixel of the image, for every $m,n\in [M]$. The Gaussian windows $\{\phi_m\}_{m=1}^M$, \editr{specified in~\eqref{eq:gauss window}}, form a $\mathcal{C}$-system on $\I$ for arbitrary $T'\subset \I$, see for instance~\cite[Example 5]{karlin1966tchebycheff}. Therefore, in \editr{view} of  Proposition~\ref{prop:noisefree}, the measure $x$ is the unique solution of Program~\eqref{eq:feas} with $\delta'=0$, \editr{provided that} $M\ge 2K+1$. This simple argument should be contrasted with the elaborate proofs of earlier 1-D results, for example Theorem~1.3 in~\cite{schiebinger2015superresolution}.

In the presence of noise, i.e., when $\delta\ge 0$, Lemma~23 in~\cite{eftekhari2018sparse} establishes that all the families of functions in Theorem~\ref{thm:main noisy simplified} are indeed $\mathcal{C}^*_{K,\varepsilon}$-systems on $\I$, provided that the endpoints of $\I$ are included in the sampling points $T'$. 
 In fact, Section 1.2 in \cite{eftekhari2018sparse} goes further and also evaluates the factors involved in the error bound for 1-D super-resolution, although arguably the result is suboptimal and there is room for improvement. In principle, those results could be in turn used to evaluate $c_1,c_2,c_3$ in the error bound of Theorem~\ref{thm:main noisy simplified}, a direction which is not pursued here. The conclusion of this section is recorded below.
\begin{cor}[Gaussian window]
Consider a nonnegative measure $x$ supported on $\I^2$. Consider also a noise level $\delta\ge 0$ and \editr{the family of} measurement functions defined in \eqref{eq:gauss window}.  For an integer $K$ and $\varepsilon\in (0,1/2]$, 
let $x_{K,\varepsilon}$ be a  $K$-sparse and $\varepsilon$-separated  nonnegative measure on $\I^2$ that approximates $x$  in the sense of~\eqref{eq:residual}. 
With $M\ge2K+2$, let~$\widehat{x}$ be a solution of Program~\eqref{eq:feas} with 
$$
\delta'\ge \l(1 + \sqrt{\frac{2M}{\sigma^2 e}} R(x,k,\varepsilon)\r)\delta, 
$$
see~\eqref{eq:residual}.  Then,
% it holds that \tb{remove it holds that}
\begin{align}
 d_{\G}\left( x,\widehat{x}\right) 
& \le c_1  \delta+ \editr{c_2(\varepsilon)} + c_3  R(x,K,\varepsilon),
\label{eq:EMD simplified gauss}
\end{align}
where $d_{\G}$ is the generalized Wasserstein metric in \eqref{eq:def of gen EMD}.  \editr{Above,} $c_1,c_2(\varepsilon),c_3$ are specified explicitly in~\eqref{eq:c3}, and depend on the true measure $x$, the separation $\varepsilon$, \editr{and the sampling locations $T'=\{t'_m\}_{m=1}^M$ in~\eqref{eq:gauss window}. In particular, it holds that $c_2(0)=0$.}
%the vectors formed by the coefficients of two polynomials constructed explicitly in the proof
\end{cor}

\section{Related Work\label{sec:related}}

The current wave of super-resolution research using convex optimization began with the two seminal papers of Cand\`{e}s and Fernandez-\editr{Granda}~\cite{Candes2014,Candes2013}. In those papers, the authors showed that a convex program with a sparse-promoting regularizer stably recovers a complex atomic measure from the low-end of its spectrum. This holds true if the minimal separation between any two spikes is inversely proportional to the maximal measured frequency, i.e., the ``bandwidth'' of the sensing mechanism.  
Many papers extended this fundamental result  to randomized models~\cite{tang2013compressed}, support recovery analysis~\cite{li2016approximate,duval2015exact,Azais2015,Fernandez-Granda2013,bendory2016stable}, denoising schemes~\cite{bhaskar2013atomic,tang2015near}, different geometries~\cite{bendory2015exact,bendory2015super,bendory2014exact,filbir2016exact,dossal2017sampling}, and incorporating prior information~\cite{mishra2015spectral}. Most of these works easily generalize to multi-dimensional signals. In addition, a special attention to multi-dimensional signals was   given in a variety of papers; see for instance~\cite{de2017exact,xu2014precise}.

The separation condition above is unnecessary for nonnegative measures, and this is the important regime on which this paper and most of this review focuses. There are a number of works that study nonnegative sparse super-resolution for atomic measures supported on a grid.
In~\cite{morgenshtern2014stable,bendory2017robust}, it was shown that for such 1-D or 2-D signals, stable reconstruction is possible without imposing a separation condition, \edit{but instead requiring a milder condition on the density of the impulses. In particular, the error grows exponentially fast as the density of the spikes increases.} 
A similar result was derived for signals on the sphere~\cite{bendory2015recovery}. 

In this paper, we focus on the grid-free setting~\editr{\cite{eftekhari2013matched,eftekhari2011probabilistic}} in which the nonnegative measure is not necessarily supported {on} a predefined grid. This is the most general regime and requires more advanced machinery and algorithms. In~\cite{schiebinger2015superresolution}, it was shown that in the absence of noise, a convex program with TV regularizer can recover--without imposing any separation--an atomic measure on the real line~\cite{schiebinger2015superresolution}. The same holds on other geometries as well~\cite[Section 5]{bendory2015exact}. However, all these results  have no stability guarantees, assume a differentiable point spread function  and make use of a TV regularizer to promote sparsity. Our Proposition~\ref{prop:noisefree} and Theorem~\ref{thm:main noisy simplified} address all these shortcomings and solve the  sparse (grid-free) image super-resolution problem in its most general form. The leap from {the 1-D results of~\cite{eftekhari2018sparse}} to 2-D} requires new techniques since the key technical ingredient, i.e., $\mathcal{C}$-systems, does not \editr{naturally} extend to higher dimensions. 

Let us add that the low-noise regime for positive 1-D super-resolution was studied in~\cite{denoyelle2017support}. 
There, it was shown that a convex program with a sparse-promoting regularizer results in the same number of spikes as the original measure when \editr{the} noise \editr{level} is small. Furthermore, the solution converges to the underlying positive measure if the signal-to-noise ratio scales like $\mathcal{O}(1/\text{sep}^2)$, where $\text{sep}$ is the minimal separation between adjacent spikes. 
In contrast to our work, the framework of~\cite{denoyelle2017support} builds upon smooth convolution kernel and uses a sparse-promoting regularizer, rather the feasibility problem considered in Program~\eqref{eq:feas}. In~\cite{poon2017multi}, it was shown that the 2-D version of the same program enjoys  similar properties 
for a pair of spikes.

Going back to  signed measures, another line of work is based on various generalizations of Prony's method~\cite{stoica2005spectral}, which encodes the support of the measure as zeros of a designed polynomial.
Such generalizations include methods like  MUSIC~\cite{schmidt1986multiple}, Matrix Pencil~\cite{hua1990matrix}, ESPRIT~\cite{roy1989esprit}, to name a few. In 1-D and in the absence of noise, these methods are guaranteed to achieve exact recovery for a complex measure without enforcing any separation. This is not true for convex programs in which separation is a necessary condition, see~\cite{tang2015resolution}.
The separation is not necessary for convex programs only for nonnegative measures, like the model considered in this paper. %Nonnegativity is not directly exploited by the Prony-like algorithms.
Stability analysis of some of these methods, under a separation condition, is found in~\cite{liao2016music,fannjiang2016compressive,moitra2015super,liao2015music,eftekhari2015greed,eftekhari2013greed}. However, their extension to 2-D is not trivial and accordingly different methods were proposed~\cite{sacchini1993two,ongie2016off,peter2017reconstruction,kunis2016multivariate,andersson2017espirit}. To the best of our knowledge, the stability of these algorithms for two or higher dimensions is not understood. 
That being said, we do not claim that convex programs are numerically superior over the Prony-like techniques and we leave comprehensive numerical study for future research.

\section{Theory} \label{sec:theory}

\subsection{Notation\label{sec:notation}}
At the risk of being redundant, let us collect here some of the notation used throughout this paper. For positive $\varepsilon$ and $T=\{t_k\}_{k=1}^K\subset \I$, let us define the neighbourhoods
\begin{equation*}
t_{k,\varepsilon} := \{t\in \I: |t-t_k| \le \varepsilon \} \subset \I,
\end{equation*} 
\begin{equation}
T_{\varepsilon} := \bigcup_{k=1}^K t_{k,\varepsilon},  
\label{eq:neigh t}
\end{equation}
and let $t_{k,\varepsilon}^C$ and $T_{\varepsilon}^C$ be the complements of these sets with respect to $\I$. Let also $\mbox{sep}(T)$ denote the minimum separation of $T$, i.e., the largest number $\nu$ for which 
\begin{equation*}
\nu\le |t_k - t_l|, \qquad k\ne l,\,\,k,l\in [K],
\end{equation*}
\begin{equation}
\nu \le |t_k-0|,\qquad \nu \le |t_k-1|.
\label{eq:sep 1d}
\end{equation}
Likewise, for positive $\varepsilon$ and $\Theta\subset\{\theta_k\}_{k=1}^K=\{(t_k,s_k)\}_{k=1}^K\subset \I^2$, we define the neighbourhoods 
\begin{equation*}
\theta_{k,\varepsilon} := 
t_{k,\varepsilon} \times s_{k,\varepsilon}
=
\{\theta\in \I^2:\|\theta-\theta_k\|_{\infty} \le \varepsilon\} 
\subset \I^2,
\end{equation*}
\begin{equation}
\Theta_{\varepsilon} := \bigcup_{k=1}^K \theta_{k,\varepsilon}
\subseteq T_\varepsilon \times S_\varepsilon
,
\label{eq:neighborhoods theta}
\end{equation}
and let $\theta_{k,\varepsilon}^C$ and $\Theta_{\varepsilon}^C$ be the complements of these sets with respect to $\I^2$.  Above, $\|\theta\|_{\infty}=\max[|t|,|s|]$ \editr{is the $\ell_\infty$-norm of} $\theta=(t,s)$.  Similarly, define the minimum separation of $\Theta$, i.e., the smallest number $\nu$ for which both \eqref{eq:sep 1d} holds and  
\begin{equation*}
\nu\le |s_k - s_l|, \qquad k\ne l,\,\,k,l\in [K],
\end{equation*}
\begin{equation}
\nu \le |s_k-0|,\qquad \nu \le |s_k-1|.
\label{eq:sep 2d}
\end{equation}

\subsection{Proof of Proposition \ref{prop:noisefree} (Sparse Measure Without Noise)} \label{sec:nonoise proof}

The following standard result is an immediate extension of \cite[Lemma 9]{eftekhari2018sparse} and, roughly speaking, states that Program~\eqref{eq:feas} is successful if a certain dual certificate $Q$ exist.
\begin{lem}
\label{lem:dual cert required}
Let $x$ be a  $K$-sparse nonnegative atomic measure supported on $\Theta\subset \operatorname{interior}(\I^2)$, see~\eqref{eq:atomic}.  Then $x$ is the unique solution of Program~\eqref{eq:feas} with $\delta'=0$ if 
\begin{itemize}
\item the  $M^{2}\times K$
matrix 
$
\left[\phi_{m}(t_{k})\phi_{n}(s_{k})\right]_{m,n,k=1}^{m=M,n=M,k=K}
$
has full column rank, and  
\item there exist  real coefficients $\{b_{m,n}\}_{m,n=1}^M$ and  polynomial  $Q(t,s)=\sum_{m,n=1}^M b_{m,n}\phi_m(t)\phi_n(s)$ such that   $Q$ is nonnegative on $\operatorname{interior}(\I^2)$ and vanishes only on $\Theta$. 
\end{itemize}
\end{lem}
The following result, proved in Appendix~\ref{sec:Proof-of-Proposition existence of dual},
states that the dual certificate required in Lemma~\ref{lem:dual cert required} exists if the number of measurements $M$ is large enough and the
measurement functions $\{\phi_{m}\}_{m=1}^M$ form a $\mathcal{C}$-system on
$\I$.

\begin{lem}[Sparse measure without noise]
\label{prop:existence of dual cert}Let $x$ be a  $K$-sparse nonnegative atomic measure supported on $\operatorname{interior}(\I^2)$. For $M\ge2K+1$, suppose that
$\{\phi_{m}\}_{m=1}^M$ form a $\mathcal{C}$-system on $\I$.
  Then, the dual certificate $Q$ prescribed in Lemma \ref{lem:dual cert required} exists.
\end{lem}

Combining Lemmas~\ref{lem:dual cert required} and~\ref{prop:existence of dual cert} completes the proof of  Proposition \ref{prop:noisefree}.

\begin{rem}[Proof technique of Lemma~\ref{prop:existence of dual cert}]
{The technical  \editr{challenge in the proof of Lemma~\ref{prop:existence of dual cert} is that}  $\mathcal{C}$-systems \editr{do not  generalize} to two dimensions.} \edita{Indeed,} to prove our claim, we effectively reduce the construction
of \editr{the dual} polynomial $Q(\theta)$ with $\theta=(t,s)$ into the construction of a number of univariate polynomials in $t$ and $s$. The key observation \editr{of the proof} is
the following. {\color{black}Suppose for simplicity that $\phi_1 \equiv 1$.} 
Recall that $\Theta = \{\theta_k\}_{k=1}^K=\{(t_k,s_k)\}_{k=1}^K$ are the impulse locations and let $T=\{t_{k}\}_{k=1}^K$, $S=\{s_{k}\}_{k=1}^K$ for short,  so that $\Theta\subseteq (T\times S)$.  Suppose that a univariate
polynomial $q_{T}(t)$ {\color{black}of $\{\phi_m(t)\}_{m=1}^M$} is nonnegative on $\I$ and only vanishes on $T$.
Similarly, consider a polynomial $q_{S}(s)$ {\color{black}of $\{\phi_n(s)\}_{n=1}^M$} that is nonnegative on $\I$ and only
vanishes on $S$. Then, the polynomial {\color{black}$Q(\theta)=q_{T}(t)+q_{S}(s)$} is nonnegative
on $\I^{2}$ and vanishes only on $T\times S$. 

\vspace{2pt}
\editr{{In general} $\Theta\subset  (T\times S)$ and, consequently, the above $Q$ will have unwanted zeros on $(T\times S)\backslash \Theta$. 
% \edita{In turn, these additional zeros lead to a suboptimal number of measurements, requiring $M=O(K^2)$ measurements to successfully recover the true measure.}
This issue may be addressed by replacing the $M^2\times K$ matrix in Lemma~\ref{lem:dual cert required} with a larger  matrix of size $M^2\times K^2$. Alternatively, we use in the proof a more nuanced argument to construct a different polynomial $Q$ that vanishes exactly on $\Theta$, without the unwanted zeros on $(T\times S) \backslash \Theta$,
%  \edita{thus requiring an optimal $M=O(K)$ number of measurements (up to a constant factor).}
see Appendix~\ref{sec:Proof-of-Proposition existence of dual} for the details. This more nuanced argument also lends itself naturally to  noisy super-resolution, as we will see later.
% and also to the case where the measurement functions along the two dimensions are different. 
} 
\end{rem}

\subsection{Geometric Intuition for Proposition \ref{prop:noisefree} \label{sec:geometry}}

Proposition~\ref{prop:noisefree} states that the imaging operator $\Phi$ in~\eqref{eq:meas model} is injective on all $K$-sparse nonnegative measures (such as $x$) provided that we take enough observations ($M\ge 2K+1$) and the measurement functions $\{\phi_m\}_{m=1}^M$ form a $\mathcal{C}$-system on $\I$. Here, we provide some geometric intuition \editr{about} the role of the dual certificate \editr{in the proof of Proposition~\ref{prop:noisefree}}. 
% We mention that the dual certificate is derived using a variety of alternative arguments in the compressed sensing and super-resolution literature~\cite{Candes2006c,Candes2014}}.

\begin{figure}[H]
\begin{center}
\subfloat[\label{fig:cone2}]{\protect\includegraphics[width=0.3\textwidth]{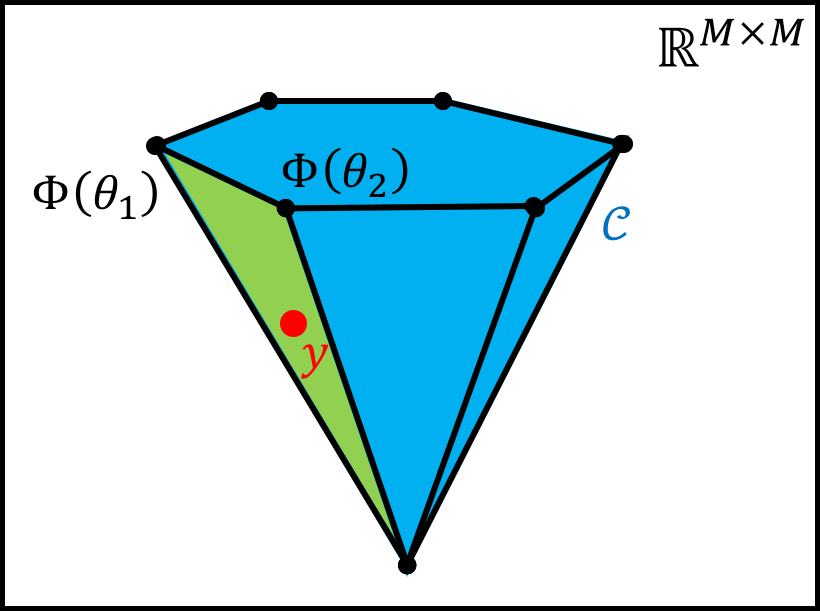}
}

\subfloat[\label{fig:hyperplane}]{\protect\includegraphics[width=0.3\textwidth]{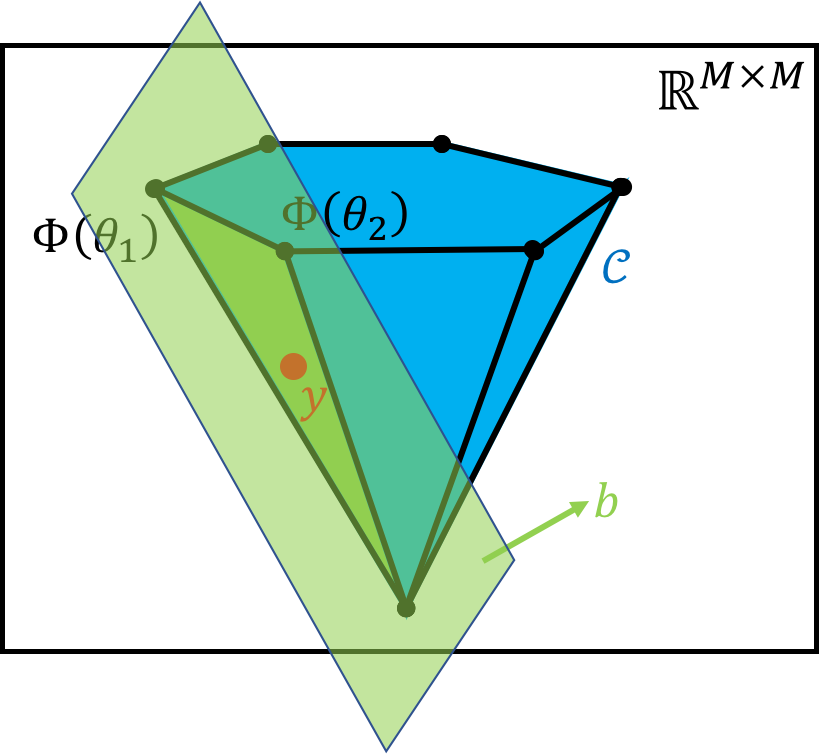}
}

\caption{This figure complements Section \ref{sec:geometry}. Figure \ref{fig:cone2} shows the conic hull $\mathcal{C}$ of the \editr{range} of the imaging operator $\Phi:\I^2\rightarrow\mathbb{R}^{M\times M}$. Note that the image $y$ \editr{in~\eqref{eq:noiseFreeMeas}} belongs to the cone $\mathcal{C}$, see \eqref{eq:yIsInC}. For  Program \eqref{eq:feas} to successfully recover the true measure $x$  from \editr{the} image $y$, it  \editr{suffices} that $\{\Phi(\theta_k)\}_{k=1}^K$ form a $K$-dimensional \editr{exposed} face of the cone $\mathcal{C}$, where $\{\theta_k\}_{k=1}^K$ is the support of \editr{the measure} $x$. This face is shown in green. In words, this condition is \edita{sufficient} for $x$ to be the unique solution of Program \eqref{eq:feas}.  
Equivalently, it \editr{suffices} to find a hyperplane, with normal vector $b$, that strictly supports $\mathcal{C}$ on this face. This \editr{latter} condition can be interpreted as finding a nonnegative polynomial of $\{\phi_m(t)\phi_n(s)\}_{m,n=1}^M$ with zeros exactly on the support  of \editr{the measure} $x$. 
  \label{fig:geometry}} 
\end{center}
\end{figure}

Let us denote $\theta=(t,s)$ for short and consider the \editr{closure of the} {conic hull} of the {dictionary} $\{\Phi(\theta)\}_{\theta\in \I^2}$ defined as
\begin{equation}
\label{eq:cone}
\mathcal{C} := \l\{\int_{\I^2} \Phi(\theta) \chi(d\theta) \,:\, \chi \mbox{ is a nonnegative measure on } \I^2\r\} \subset \mathbb{R}^{M\times M}.
%
%\sum_{l=1}^{k'} \alpha_l \cdot \Phi(\tau_l)\,:\, k' \mbox{ is finite, } \{\alpha_l\}_l\mbox{ are positive, and } \{\tau_l\}_l\subset I \r\} \subset \mathbb{R}^m.
\end{equation}
By the continuity of $\Phi$ and with an application of the dominated convergence theorem, it is easy to verify that $\mathcal{C}$ is a closed convex cone, i.e., $\mathcal{C}$ is a homogeneous and closed convex subset of $\mathbb{R}^{M\times M}$. 
When $\{\phi_m\}_{m=1}^M$ form a $\mathcal{C}$-system on $\I$, it also not difficult to verify that $\{\Phi(t_l,s_l)\}_{l=1}^{M^2}$ are linearly independent matrices in $\mathbb{R}^{M\times M}$. 
%for any choice of distinct $\{t_l\}_{l=1}^M\subset \I$ and distinct $\{s_l\}_{l=1}^M\subset \I$. 
This in particular implies that $\mathcal{C}$ is a convex body, i.e., the interior of $\mathcal{C}$ is not empty. Note also that $y\in \mathcal{C}$ because 
\begin{equation}
y=\int_{\I^2} \Phi(\theta) x(d\theta) = \sum_{k=1}^M a_k \Phi(\theta_k).
\label{eq:yIsInC}
\end{equation}
For Program~\eqref{eq:feas} to successfully recover \editr{the measure} $x$, it \editr{suffices} that 
\begin{equation}
\mathcal{A} = \mbox{cone}\l(\{\Phi(\theta_k)\}_{k=1}^K
\r) 
= \l\{ \sum_{k=1}^K \alpha _k \Phi(\theta_k)\,:\, \alpha_k \ge 0,\,\forall k\in [K] \r\},
\label{eq:def of face}
\end{equation}
is a $K$-dimensional {exposed face} of the cone $\mathcal{C}$. \edita{See \cite[\textsection 18]{rockafellar1970convex} for the definition of exposed face.} This \editr{in turn} happens if and only if we can find a hyperplane with normal vector $b\in\mathbb{R}^{M\times M}$  that {strictly supports} the cone  $\mathcal{C}$ at $\mathcal{A}$, i.e., when we can find $b$ such that 
\begin{equation}
\begin{cases}
\l\langle b,c\r\rangle =0, & \forall c\in \mathcal{A},\\
\l\langle b,c\r\rangle >0, & \forall c\in \mathcal{C}\backslash \mathcal{A}.
\end{cases}
\label{eq:face cnds}
\end{equation}
Invoking \eqref{eq:def of face}, we find that \eqref{eq:face cnds} is equivalent to finding $b\in \mathbb{R}^{M\times M}$ such that 
\begin{equation}
\begin{cases}
\l\langle b,\Phi(\theta_k) \r\rangle =0, &  k\in [K],\\
\l\langle b,\Phi(\theta)\r\rangle >0, &  \theta \notin \{\theta_k\}_{k=1}^K.
\end{cases}
\end{equation}
In other words, \editr{for Program~\eqref{eq:feas} to successfully recover the measure $x$}, it \editr{suffices} to find a ``polynomial'' 
$$
Q(\theta)=Q(t,s):= \langle b,\Phi(\theta)\rangle = \sum_{m,n=1}^{M} b_{m,n} \phi_{m}(t)\phi_n(s), 
$$
that vanishes on $\{\theta_k\}_{k=1}^K$ and is positive elsewhere on $\I^2$. Building on the results in \cite{eftekhari2018sparse}, we construct one such polynomial in Section \ref{sec:nonoise proof} when $M\ge 2K+1$. 
 It is worth noting that the {polar} of the cone $\mathcal{C}$, itself another convex cone in $\mathbb{R}^{M\times M}$,  consists of the coefficients of all nonnegative polynomials of $\{\phi_m\phi_n\}_{m,n=1}^M$ on $\I^2$ and in particular the coefficient vector $b$ above belongs to an $(M^2-K)$-dimensional face of this polar cone \cite{karlin1966tchebycheff}. {We refer the reader to Figure~\ref{fig:geometry} for an illustration of the convex geometry underlying the problem of nonnegative super-resolution.}

\subsection{Proof of Theorem~\ref{thm:main noisy simplified} (Arbitrary Measure with Noise)\label{sec:noisy}}

In this section, we will prove the main result of this paper, \editr{i.e.,} Theorem~\ref{thm:main noisy simplified}.
For an integer $K$ and $\varepsilon\in (0,1/2]$, let  $x_{K,\varepsilon}$ be a $K$-sparse and $\varepsilon$-separated nonnegative measure on $\I^2$ that approximates $x$ \editr{with residual $R(x,K,\varepsilon)$,} in the sense of~\eqref{eq:residual}. Let $\Theta=\{\theta_k\}_{k=1}^K=\{(t_k,s_k)\}_{k=1}^K \subset \interior(\I)^2$ be the support of $x_{K,\varepsilon}$, and set $T=\{t_k\}_{k=1}^K$ and $S=\{s_k\}_{k=1}^K$ for short.  Consider also the neighbourhoods $\{\theta_{k,\varepsilon}\}_{k=1}^K\subseteq\I^2$ and $\Theta_\varepsilon=\cup_{k=1}^K \theta_{k,\varepsilon}$  defined in~\eqref{eq:neighborhoods theta}.  Let 
$\{\theta_{k,\varepsilon}^{C}\}_{k=1}^K$ and $\Theta_{\varepsilon}^{C}$ denote 
the complements of these sets with respect to~$\I^{2}$. 

\editr{Before turning to the details, let us outline the proof.}
% In \editr{Sections~\ref{sec:dual certs} and~\ref{sec:existence of dual certs}}, we  bound the error $d_{\G}(x_{K,\varepsilon},\widehat{x})$ \tb{I think you can remove this sentence}. 
% \editr{More specifically,} 
We will  show in Section~\ref{sec:dual certs}  that the existence of certain
dual certificates leads to \editr{a} stable recovery of $x_{K,\varepsilon}$ with Program~\eqref{eq:feas}.
\editr{Then, we show in} Section~\ref{sec:existence of dual certs} that these certificates exist under certain conditions on the imaging apparatus. 
\editr{Finally, in Section~\ref{sec:completeMainProof}, we complete the proof of Theorem \ref{thm:main noisy simplified} by applying} the triangle inequality to control  $d_{\G}(x,\widehat{x})$ \editr{as}
$$
d_{\G}(x,\widehat{x}) \le d_{\G}(x,x_{K,\varepsilon})+d_{\G}(x_{K,\varepsilon},\widehat{x}) \le  R(x,K,\varepsilon)+ d_{\G}(x_{K,\varepsilon},\widehat{x}).
\qquad \mbox{(see \eqref{eq:residual})}
$$

\subsubsection{Dual Certificates\label{sec:dual certs}}

Lemmas~\ref{lem:dual cert noise} and~\ref{lem:dual cert noise 2}
below show that Program~\eqref{eq:feas} stably recovers \editr{the atomic measure} $x_{K,\varepsilon}$ in the
presence of noise, provided that certain dual certificates exist.
The proofs, \editr{ which appear in Appendices~\ref{sec:proof dual cert noise} and~\ref{sec:proof dual cert noise 2},} are standard and Lemmas \ref{lem:dual cert noise} and \ref{lem:dual cert noise 2} below are  extensions of, respectively, Lemmas~\editr{16 and~17} in~\cite{eftekhari2018sparse}.
% \tb{(Please make sure that this is the right numbering in the published version of [1])}. 
\editr{In particular, Lemma~\ref{lem:dual cert noise} below controls the recovery error away from the support $\Theta$ of $x_{K,\varepsilon}$, whereas
 Lemma~\ref{lem:dual cert noise 2}  controls the error near the
support. The latter features an approximate dual certificate, and shares some broad similarities with~\cite{candes2010probabilistic}.}
%\editr{In particular, even though the proof machinery is different from~\cite{candes2010probabilistic}, Lemma~\ref{lem:dual cert noise 2} features an approximate dual certificate.}
% Both results were at length discussed in~\cite{eftekhari2018sparse}. 
%In \editr{words}, Lemma~\ref{lem:dual cert noise} below controls the recovery error away from the support $\Theta$ of $x_{K,\varepsilon}$, \editr{whereas}
% Lemma~\ref{lem:dual cert noise 2} below controls the error near the
%support. 

\begin{lem}[\editr{Error away from the support}]
\label{lem:dual cert noise}
 Let $\widehat{x}$ be a solution of Program~\eqref{eq:feas} with \editr{$\delta'\ge \d$} and set $h:=\widehat{x}-x_{K,\varepsilon}$ to be the error. Fix a  positive scalar $\bar{g}$. Suppose that there exist real coefficients $\{b_{m,n}\}_{m,n=1}^{M}$ and 
a polynomial 
\begin{equation}
Q(\theta)=Q(t,s) = \sum_{m,n=1}^M b_{m,n}\phi_{m}(t)\phi_n(s),
\label{eq:QFarAway}
\end{equation}
such that
\begin{align}
Q(\theta)\ge G(\theta):=\begin{cases}
0, & \text{when there exists }k\in[K]\text{ such that }\theta\in \theta_{k,{\varepsilon}},\\
\bar{g}, & \text{elsewhere \editr{in} }\interior(\I^2),
\end{cases}
\label{eq:propsOfQFar}
\end{align}
where the equality holds on $\Theta=\{\theta_k\}_{k=1}^K$. Then \editr{it holds that }
\begin{equation}
\int_{\Theta_{\varepsilon}^{C}} h(d\theta)\le2\|b\|_{\mathrm{F}}\delta'/\bar{g},\label{eq:o obar bound}
\end{equation}
where $b\in\mathbb{R}^{M\times M}$ is the matrix formed by the coefficients
$\{b_{m,n}\}_{m,n=1}^M$. 
\end{lem}

\begin{lem}[\editr{Error near the support}]
\label{lem:dual cert noise 2}
%  Suppose that the dual certificate $Q$ in Lemma \ref{lem:dual cert noise}
% exists. 
\editr{Let $\widehat{x}$ be a solution of Program~\eqref{eq:feas} with $\d'\ge \d$ and set $h:=\widehat{x}-x_{K,\varepsilon}$ to be the error. For 
% \tb{any?specific?} 
$\alpha\in [0,1]$,} 
% \tb{$\alpha$ can be zero or one?}
suppose  that there exist real coefficients $\{b^0_{m,n}\}_{m,n=1}^M$ and a polynomial 
\begin{equation}
Q^0(\theta)=Q^0(t,s) = \sum_{m,n=1}^M b_{m,n}^{0}\phi_{m}(t)\phi_n(s),
\label{eq:Q0Defn}
\end{equation}
 such that 
\begin{equation}
Q^0(\theta)\ge G^0(\theta):=
\begin{cases}
1, & \text{when there exists } k\in [K] \text{ such that }\theta\in \theta_{k,{\varepsilon}}\text{ and }\int_{\theta_{k,{\varepsilon}}} h(d\theta)>0,\\
-1, & \text{when there exists }k\in [K] \text{ such that }\theta\in \theta_{k,{\varepsilon}}\text{ and }\int_{\theta_{k,{\varepsilon}}} h(d\theta)\le0,\\
\editr{-1+\alpha} & \editr{\text{when there exists }} \editr{k\in [K]} \text{ such that }\theta= \theta_{k}\text{ and }\int_{\theta_{k,{\varepsilon}}} h(d\theta)\le0,\\
\editr{-1}, & \text{elsewhere in }\interior(\I^2),
\end{cases}\label{eq:q0}
\end{equation}
where the equality holds on $\Theta$. Then \editr{it holds that}
\editr{
\begin{equation}
 \sum_{k=1}^K \left|\int_{\theta_{k,{\varepsilon}}} h(d\theta)\right|\le \alpha\|x_{K,\varepsilon}\|_{\mathrm{TV}}+ 2\l(\|b^0\|_{\mathrm{F}}+ \frac{\|b\|_\F}{\overline{g}} \r)\delta',\label{eq:overall neighborhoods}
\end{equation}
}
where $b^{0}\in\mathbb{R}^{M\times M}$ is
the matrix formed by the coefficients $\{b_{m,n}^{0}\}_{m,n=1}^M$, \editr{and the matrix $b$ was introduced in Lemma~\ref{lem:dual cert noise}.}
\end{lem}

By combining Lemmas~\ref{lem:dual cert noise} and~\ref{lem:dual cert noise 2}, the next result bounds the error $d_{\G}(x_{K,\varepsilon},\widehat{x})$. The proof  is omitted as \editr{the steps are}   identical to \editr{those taken in} Lemma 18 in~\cite{eftekhari2018sparse}. 
\begin{lem}[Error in Wassertein metric]
\label{lem:EMD}  Suppose that the dual certificates $Q$ and $Q^0$ in \edita{Lemmas}~\ref{lem:dual cert noise}
and~\ref{lem:dual cert noise 2} exist. Then,
\begin{equation}
d_{\G}\left( x_{K,\varepsilon},\widehat{x}\right)\le
\left( \editr{\frac{8\|b\|_{\mathrm{F}}}{\bar{g}}}+6\|b^{0}\|_{\mathrm{F}}\right)\delta'+\editr{\l(\varepsilon + 3 \alpha\r)} \| x_{K,\varepsilon} \|_{\mathrm{TV}}.
\label{eq:EMD}
\end{equation}
\end{lem}
In order to apply {Lemma}~\ref{lem:EMD}, we must first show that the dual certificates $Q$ and $Q^0$ specified in {Lemmas}~\ref{lem:dual cert noise} and~\ref{lem:dual cert noise 2} exist. In the next section, we construct these certificates under certain conditions on the imaging apparatus.  

\subsubsection{Existence of the Dual Certificates\label{sec:existence of dual certs}}

To \editr{construct} the dual certificates required in {Lemmas}~\ref{lem:dual cert noise}
and~\ref{lem:dual cert noise 2}, some preparation is necessary.
\editr{For notational convenience, throughout we model $x_{K,\varepsilon}$ by~\eqref{eq:atomic}, with $x$ therein replaced with $x_{K,\varepsilon}$.}
\editr{Then} recall \editr{from Section~\ref{sec:notation}} that $\Theta=\{\theta_{k}\}_{k=1}^K=\{(t_{k},s_{k})\}_{k=1}^K$ are the impulse locations \editr{of $x_{K,\varepsilon}$} and  let $T=\{t_{k}\}_{k=1}^K$ and $S=\{s_{k}\}_{k=1}^K$ for short.
In particular, \editr{note that} $\Theta\subseteq T\times S$.

For an index set $\Omega\subseteq [K]$  and its complement $[K]\backslash\Omega$, we set $T_{\Omega}=\{t_{k}\}_{k\in\Omega}$
and $S_{[K]\backslash\Omega}=\{s_{k}\}_{k\in[K]\backslash\Omega}$ \editr{for brevity}. 
For a finite set \editr{of distinct points} $T'\subset\I$ and  \editr{$\varepsilon\in (0,\text{sep}(T')]$}, let us \editr{also} define the function \editr{$F_{T'}:\I\rightarrow\mathbb{R}$ as}
\begin{equation}
F_{T'}(t):=\begin{cases}
0, & \text{when there exists }t'\in T'\text{ such that }t\in t'_{\varepsilon},\\
\pline, & \text{elsewhere on }\I,
\end{cases}\label{eq:P_T'}
\end{equation}
where $t'_{\varepsilon}=\left\{ t\in\I:|t-t'|\le\varepsilon\right\} $ \editr{is the $\varepsilon$-neighborhood of $t'$.}
The following result, proved in Appendix~\ref{sec:Proof-of-Proposition existence of 1st dual},
 states that the dual certificate \editr{prescribed} in {Lemma}~\ref{lem:dual cert noise} exists if both $\{F_{T_{\Omega}}\}\cup\{\phi_{m}\}_{m=1}^M$ and $\{F_{S_{\Omega}}\}\cup\{\phi_{m}\}_{m=1}^M$ are $\mathcal{C}^*$-systems for \editr{every} index set $\Omega\subseteq[K]$.

\begin{prop}[\editr{Dual certificate for faraway error}]
\label{prop:existence of 1st dual}  Suppose that
$\{\phi_{m}\}_{m=1}^M$ form a $\mathcal{C}$-system on $\I$ with $M\ge2K+2$.
For every index set $\Omega\subseteq[K]$, suppose also that $\{F_{T_{\Omega}}\}\cup\{\phi_{m}\}_{m=1}^M$
and $\{F_{S_{\Omega}}\}\cup\{\phi_{m}\}_{m=1}^M$ are
both $\mathcal{C}^*_{K,\varepsilon}$-systems on $\I$, see~\eqref{eq:P_T'}. Then the dual certificate $Q$, \editr{as} specified in 
{Lemma}~\ref{lem:dual cert noise}, exists with
\begin{equation}
\oline=2^{K-2}.\label{eq:o obar vs p pbar}
\end{equation}
% where $\bar{g}$ is the constant from  {Lemma}~\ref{lem:dual cert noise}.
\end{prop}

\editr{Next,} for  \editr{$t',s'\in \I$} and \editr{$\varepsilon\in(0,1/2]$}, let us define the functions \editr{ $F_{t'}^{\pm}:\I \rightarrow \mathbb{R}$ and $F_{s'}^{+ }: \I \rightarrow \mathbb{R}$ as}
\editr{\[
F_{t'}^{\pm}(t):=
\begin{cases}
\pm 1, & \text{when }t\in t'_{\varepsilon},\\
% %\text{ and }\int_{\theta{}_{k,\varepsilon}}h(d\theta) \ge 0,\\ %\lesseqgtr
% %-1, & \text{when }t\in t_{k,\varepsilon}\text{ and }\int_{\theta{}_{k,\varepsilon}}h(d\theta) < 0,\\
 0, & \text{elsewhere on }\I,
\end{cases}
\]
\begin{equation}
F_{s'}^{+}(s):=\begin{cases}
 1, & \text{when }s\in s'_{\varepsilon},\\
%-g\left(s-s_{l}\right), & \text{when there exists }l\ne k\text{ such that }s\in s_{l,\varepsilon},\\
0, & \text{elsewhere on }\I,
\end{cases}
\label{eq:GkTnS}
\end{equation}
where $t'_\varepsilon$ and $s'_\varepsilon$ are the $\varepsilon$-neighborhoods of $t'$ and $s'$, respectively.}
The following result, proved in Appendix~\ref{sec:Proof-of-Proposition existence of 2nd dual},
 states that the dual certificate, \editr{prescribed} in 
\edita{Lemma}~\ref{lem:dual cert noise 2}, exists when certain $\mathcal{C}^*$-systems exist. 

\begin{prop}[\editr{Dual certificate for nearby error}]
\label{prop:existence of 2nd dual} For $M\ge2K+2$,
suppose that $\{\phi_{m}\}_{m=1}^M$ form a $\mathcal{C}$-system on $\I$. 
 For every $k\in [K]$, suppose also that $\{F_{t_k}^+\}\cup\{\phi_{m}\}_{m=1}^M$, $\{F_{t_k}^{-}\}\cup\{\phi_{m}\}_{m=1}^M$ ,  and $\{F_{s_k}^{+}\}\cup\{\phi_{m}\}_{m=1}^M$ are all $\mathcal{C}^*_{K,\varepsilon}$-systems on $\I$, see \eqref{eq:GkTnS}. Then the dual certificate $Q^{0}$, \editr{as specified}
in Lemma~\ref{lem:dual cert noise 2}, exists \editr{with $\alpha=\alpha(\varepsilon)= 1- 1/q_{\max}^{\pi_*}(\varepsilon)$, where $q_{\max}^{\pi_*}(\varepsilon)$ is defined \editr{in~\eqref{eq:qMax} and~\eqref{eq:alphaDefined}.} In particular, $ \alpha(0) = 0$.}  
\end{prop}

\subsubsection{Completing the Proof of Theorem \ref{thm:main noisy simplified} \label{sec:completeMainProof}}

Recall that the imaging apparatus is $L$-Lipschitz \editr{in the sense of}~\eqref{eq:Lipschitz assumption}. 
\editr{From this assumption and} the triangle inequality, it follows that 
\begin{align}
\l\|y-\int_{\I^2} \Phi(\theta) \, x_{K,\varepsilon} (\der \theta)\r\|_{\mathrm{F}} & \le \l\|y-\int_{\I^2}  \Phi(\theta) \, x(\der \theta)\r\|_{\mathrm{F}}+ \l\|\int_{\I^2}  \Phi(\theta)\,  (x(d\theta)-x_{K,\varepsilon} (\der \theta)) \r\|_{\mathrm{F}} \nonumber \\
& \le \delta+ L\cdot d_{\G}\left(x,x_{K,\varepsilon} \right) 
\qquad \mbox{(see (\ref{eq:meas model},\ref{eq:Lipschitz assumption}))} \nonumber\\
& = \delta + L\cdot R(x,K,\varepsilon) := \delta'.
\qquad \mbox{(see \eqref{eq:residual})}
%& \le \delta + L\varepsilon \|x\|_{\mathrm{TV}} =: \delta'.
%
\label{eq:def of deltap}
\end{align}
\editr{In words}, a solution $\widehat{x}$ of Program~\eqref{eq:feas}, with $\delta'$ specified above, can be \editr{thought of} as an estimate of $x_{K,\varepsilon}$. \editr{Recall that} we also constructed the \editr{prescribed} dual certificates \editr{$Q$ and $Q^0$} in Section~\ref{sec:existence of dual certs}, \editr{see} Propositions~\ref{prop:existence of 1st dual} and~\ref{prop:existence of 2nd dual}. Consequently, Lemmas \ref{lem:dual cert noise}\editr{-\ref{lem:EMD} are in force}. The following argument \editr{thus} completes the proof of Theorem~\ref{thm:main noisy simplified}:
\begin{align}
d_{\G}(x,\widehat{x})
& \le d_{\G}(x,x_{K,\varepsilon}) + d_{\G}(x_{K,\varepsilon},\widehat{x}) 
\qquad \mbox{(triangle inequality)}
\nonumber\\
& \le R(x,K,\varepsilon)+ \left( \editr{\frac{8\|b\|_{\mathrm{F}}}{\bar{g}}}+6\|b^{0}\|_{\mathrm{F}}\right)\delta'+\editr{\l(\varepsilon + 3 \alpha(\varepsilon)\r)} \| x_{K,\varepsilon} \|_{\mathrm{TV}}.
\qquad \mbox{(see~(\ref{eq:residual},\ref{eq:EMD}))}
\nonumber\\
& =   R(x,K,\varepsilon)+ \l(\editr{\frac{8\|b\|_{\mathrm{F}}}{\bar{g}}}+6\|b^{0}\|_{\mathrm{F}}\r)
(\delta+L\cdot R(x,K,\varepsilon))+\editr{(\varepsilon+3\alpha(\varepsilon))}\| x_{K,\varepsilon} \|_{\mathrm{TV}} 
\qquad \mbox{(see~\eqref{eq:def of deltap})} \nonumber\\
& = \l( \editr{\frac{8\|b\|_{\mathrm{F}}}{\oline}} +6\|b^{0}\|_{\mathrm{F}} \r) \delta+ \l( \editr{\frac{8L\|b\|_{\mathrm{F}}}{\oline} } +6L\|b^0\|_{\mathrm{F}}+1  \r)R(x,K,\varepsilon) + \editr{(\varepsilon+3\alpha(\varepsilon))}\|x_{K,\varepsilon}\|_{\mathrm{TV}} \nonumber\\
& = \l( \editr{2^{5-K}} \|b\|_{\mathrm{F}}+6\|b^{0}\|_{\mathrm{F}} \r) \delta+ \l( \editr{2^{5-K}} L\|b\|_{\mathrm{F}}+6L\|b^0\|_{\mathrm{F}}+1  \r)R(x,K,\varepsilon) \nonumber\\
& \qquad\qquad 
+ \editr{(\varepsilon+3\alpha(\varepsilon))}\|x_{K,\varepsilon}\|_{\mathrm{TV}}.
\qquad \text{(see~\eqref{eq:o obar vs p pbar})}
\label{eq:pre-ult-noisy-proof}
\end{align}
\editr{The first inequality above holds because the generalized Wasserstein distance $d_{\G}$ in~\eqref{eq:def of gen EMD} indeed satisfies the triangle inequality}, see Proposition~5 in \cite{piccoli2012generalized}. 
\edita{In the last line above, it would be more convenient to relate $\|x_{K,\epsilon}\|_{\mathrm{TV}}$ back to $\|x\|_{\mathrm{TV}}$.  \editr{To that end, we} write that 
\begin{align}
\|x_{K,\epsilon}\|_{\mathrm{TV}} & = d_{\G}(x_{K,\epsilon},0) 
\qquad \text{(see \eqref{eq:def of gen EMD})}
\nonumber\\
& \le d_{\G}(x,x_{K,\epsilon} ) + d_{\G}(x,0)
\qquad \text{(triangle inequality)} \nonumber\\
& =  d_{\G}(x,x_{K,\epsilon} ) + \|x\|_{\mathrm{TV}}
\qquad \text{(see \eqref{eq:def of gen EMD})} \nonumber\\
& \le d_{\G}(x,0) + \|x\|_{\mathrm{TV}}
\qquad \text{(see \eqref{eq:residual})} \nonumber\\
& = 2 \|x\|_{\mathrm{TV}}. 
\qquad \text{(see \eqref{eq:def of gen EMD})}
\label{eq:xke-to-x-final}
\end{align}
}
{Finally, \edita{in light of (\ref{eq:pre-ult-noisy-proof},\ref{eq:xke-to-x-final})}, the \editr{components of error bound} in Theorem~\ref{thm:main noisy simplified} are given explicitly \editr{as}:
\begin{align}
c_1 &= \editr{32} \|b\|_{\mathrm{F}}+6\|b^{0}\|_{\mathrm{F}}, \nonumber\\
c_2\editr{(\varepsilon)} &= \editr{(\varepsilon+3\alpha(\varepsilon))} \|x\|_{\mathrm{TV}}, \nonumber\\
c_3 &= \editr{32} L\|b\|_{\mathrm{F}}+6L\|b^0\|_{\mathrm{F}}+1.  \label{eq:c3}
\end{align}	
}

\section{Perspective} \label{sec:perspective}

In this paper, we have shown that a simple convex feasibilty  program is guaranteed to robustly recover a sparse (nonnegative) image in the presence of model mismatch and additive noise, under certain conditions on the imaging apparatus. No sparsity-promoting regularizer or separation condition is needed, and the techniques used here are arguably simple and intuitive. In other words, we have described when the imaging apparatus acts as an injective map  \editr{over} all sparse images and when we can stably find its inverse. These results build upon and extend a recent  \editr{paper}~\cite{eftekhari2018sparse} which focuses  on 1-D signals. The extension to images, however, requires novel constructions of interpolating polynomials, called dual certificates. In practice, many super-resolution problems appear in even higher dimensions. While we believe that similar results hold in any dimension, it is yet to be proven. Similarly, the super-resolution problem was studied in different non-Euclidean  geometries for complex measures and under a separation condition~\cite{bendory2014exact,de2015non,bendory2015exact,bendory2015super,filbir2016exact}. It would be interesting to examine whether our results, which are based on the {properties of} Chebyshev systems, extend to these non-trivial geometries and to manifolds in general.

Verifying the conditions on the window for stable recovery in Theorem~\ref{thm:main noisy simplified} is rather \editr{ponderous}. As an example, we have shown that the Gaussian window, a ubiquitous model of convolution kernels, satisfies those conditions.  It is important to identify other such admissible windows and, if possible, simplify the conditions on the window in Theorem~\ref{thm:main noisy simplified}.  Another interesting research direction is deriving the optimal separation $\varepsilon$ (as a function of noise level $\delta$) that minimizes the right-hand side of the error bound in~\eqref{eq:EMD simplified}. Such a result will provide the tightest  error bound for Program~\eqref{eq:feas}.

This work has focused solely on the theoretical performance of Program~\eqref{eq:feas}. 
It is essentially important to understand, even numerically, the pros and cons of the different localization algorithms suggested in the literature. For instance, it would be interesting to investigate whether the sparse-promoting regularizer, albeit not necessary for our analysis of nonnegative measures, reduces the recovery error.

\section*{Acknowledgements}
\editr{When preparing this manuscript,} AE \editr{was} supported by the Alan Turing Institute under the EPSRC grant EP/N510129/1 and partially by the Turing Seed Funding grant SF019. GT \editr{was} supported by the NSF grant CCF-1704204 and the DARPA Lagrange Program under ONR/SPAWAR contract N660011824020. AE would like to thank Jared Tanner and Bogdan Toader for their insightful feedback. {TB would  like to thank Amit Singer for his support and Nir Sharon for helpful discussions on Chebyshev systems. \editr{We} \editr{are deeply indebted to the anonymous  reviewers of this work for their careful and detailed comments. We would also like to sincerely thank Jean-Baptiste Seby for spotting an error in the earlier version of this manuscript. } 
% \tb{I went through the references and mads some small updates (a couple of papers were published since last time)}
}

\appendix

\section{Proof of  Lemma \ref{prop:existence of dual cert} \label{sec:Proof-of-Proposition existence of dual}}

Consider the $M^{2}\times K$ matrix 
\[
A=\left[\phi_{m}(t_{k})\phi_{n}(s_{k})\right]_{m,n,k=1}^{m=M,n=M,k=K}.
\]
Note that $A$ is a column-submatrix of the $M^{2}\times K^{2}$ matrix
\[
B=\left[\phi_{m}(t_{k})\phi_{n}(s_{l})\right]_{m,n,k,l=1}^{m=M,n=M,k=K,l=K}.
\]
Therefore, to show that $A$ has full column rank, it suffices to show that
$B$ is nonsingular. Note that $B$ itself can be written as the Kronecker product
of two $M\times K$ matrices, i.e., 
\[
B=\left[\phi_{m}(t_{k})\right]_{m,k=1}^{m=M,k=K}\varotimes\left[\phi_{n}(s_{l})\right]_{n,l=1}^{n=M,l=K}=:B_{1}\otimes B_{2},
\]
where $\otimes$ stands for Kronecker product. Since by assumption $\{\phi_{m}\}_{m=1}^M$ form a $\mathcal{C}$-system on $\I$ and $M\ge2K+1\ge K$, both $B_{1}$ and $B_{2}$ are nonsingular. It follows that $B$ too is nonsingular, as claimed.

Next we recall \edita{Lemma 15 from~\cite{eftekhari2018sparse}, originally \editr{proved in} \cite[Theorem 5.1]{karlin1966tchebycheff}}, \editr{stated} below for convenience.  

\begin{lem}[\editr{Univariate polynomial of a $\mathcal{C}$-system}]
\label{lem: 3 rep}Consider a set $T'\subset\I$ of size
$K'$. With $M\ge2K'+1$, suppose that $\{\phi_{m}\}_{m=1}^M$ form
a $\mathcal{C}$-system on $\I$. Then, there exist coefficients $\{b_m\}_{m=1}^M$ such that the polynomial $q_{T'}=\sum_{m=1}^M b_{m}\phi_{m}$ is nonnegative on $\I$ and vanishes only on $T'$. 
\end{lem}
Recall that $\Theta = \{\theta_k\}_{k=1}^K=\{(t_k,s_k)\}_{k=1}^K$ are the impulse locations, and let us set $T=\{t_{k}\}_{k=1}^K$ and $S=\{s_{l}\}_{l=1}^K$ for short.
For an index set $\Omega\subseteq[K]$, let $[K]\backslash\Omega$
denote its complement with respect to $[K]$. Let also $T_{\Omega}=\{t_{k}\}_{k\in\Omega}$
and $S_{[K]\backslash\Omega}=\{s_{k}\}_{k\in[K]\backslash\Omega}$.
By assumption, $\{\phi_{m}\}_{m=1}^M$ form a $\mathcal{C}$-system on $\I$
with $M\ge2K+1$. Therefore, for every index set $\Omega\subseteq[K]$ and \editr{in view of Lemma}~\ref{lem: 3 rep},
there exist polynomials $q_{T_{\Omega}}$
and $q_{S_{[K]\backslash\Omega}}$ that are nonnegative on $\I$ and
vanish only on $T_{\Omega}$ and $S_{[K]\backslash\Omega}$, respectively.

Let us form the polynomial 
\begin{equation}
Q(\theta)=Q(t,s)=\sum_{\Omega\subseteq[K]}q_{T_{\Omega}}(t)\cdot q_{S_{[K]\backslash\Omega}}(s),\label{eq:poly sum 0}
\end{equation}
where the sum is over all subsets of $[K]$. Evidently, $Q$ is nonnegative
on $\I^{2}$ since each summand above is nonnegative. We next verify that
$Q$ only vanishes on $\Theta$. To that end, consider $\theta_{k}=(t_{k},s_{k})\in\Theta$
with $k\in[K]$ and an index set $\Omega\subseteq[K]$. There are
two possibilities. Either $k\in\Omega$, in which case $q_{T_{\Omega}}(t_{k})=0$.
Or $k\in[K]\backslash\Omega$, in which case $q_{S_{[K]\backslash\Omega}}(s_{k})=0$.
In both cases, the product vanishes, i.e., $q_{T_{\Omega}}(t_{k})\cdot q_{S_{[K]\backslash\Omega}}(s_{k})=0$.
Since the choice of $\Omega$ was arbitrary, it follows from~\eqref{eq:poly sum 0}
that $Q(\theta_{k})=Q(t_{k},s_{k})=0$ for every $k\in[K]$. 

On the other hand, suppose that $\theta\in\Theta^{C}$. The first
possibility is that $\theta=(t,s)\in T^{C}\times S^{C}\subseteq\Theta^{C}$,
i.e., $t\in T^{C}$ and $s\in S^{C}$. For arbitrary index set $\Omega\subseteq[K]$, note that 
$q_{T_{\Omega}}(t)\cdot q_{S_{[K]\backslash\Omega}}(s)>0$ by design.
It follows from~\eqref{eq:poly sum 0} that $Q(\theta)=Q(t,s)>0$
when $\theta\in T^{C}\times S^{C}$. The second possibility is that
$\theta=(t_{k},s_{l})$ with $k\ne l$ and $k,l\in[K]$. There always
exists $\Omega_{0}\subset[K]$ such that $t_{k}\in[K]\backslash\Omega_{0}$
and $s_{l}\in\Omega_{0}$. For such $\Omega_{0}$, it holds that $q_{T_{\Omega}}(t_{k})\cdot q_{S_{[K]\backslash\Omega}}(s_{l})>0$.
{For instance, one can choose $\Omega_{0}=\{s_l\}$ for which both $q_{T_{\{s_l\}}}(t_{k})$ and $q_{S_{[K]\backslash\{s_l\}}}(s_l)$ are strictly positive.} 
Consequently, $Q(\theta)>0$ by~\eqref{eq:poly sum 0} when $\theta\in\Theta^{C}\backslash(T^{C}\times S^{C})$.
In conclusion, $Q$ is positive and vanishes only on $\Theta$, as
claimed. This completes the proof of  \ref{prop:existence of dual cert}.

\section{Proof of Lemma~\ref{lem:dual cert noise}}\label{sec:proof dual cert noise}

For notational convenience, throughout we model $x_{K,\varepsilon}$  by~\eqref{eq:atomic}, with $x$ therein replaced with $x_{K,\varepsilon}$. 
By feasibility of both $\widehat{x}$ and $x_{K,\varepsilon}$ for Program~\eqref{eq:feas}, and after applying the triangle inequality, we observe that 
\begin{align}
    \l\| \int_{\I^2}\Phi(\theta)\, h(\der \theta) \r\|_{\mathrm{F}} \le 2\d',
    \label{eq:feasUsed}
\end{align}
where $\theta=(t,s)$. 
Next, the existence of the dual certificate allows us to write that 
\begin{align}
 \oline \int_{\Theta_\varepsilon^C} h(\der \theta) 
 & \le \int_{\Theta_\varepsilon^C} G(\theta)\, h(\der \theta) 
 \qquad (\text{\eqref{eq:propsOfQFar} and the error } h \text{ is nonnegative on }\Theta_{\varepsilon}^C)
\nonumber\\ 
& = \int_{\Theta_\varepsilon^C} G(\theta)\, h(\der \theta)   + \sum_{k=1}^K \int_{\theta_{k,\varepsilon}} G(\theta)\, h(\der \theta) 
\qquad \text{(see (\ref{eq:neighborhoods theta},\ref{eq:propsOfQFar}))}
 \nonumber\\
 & = \int_{\I^2} G(\theta)\, h(\der \theta) 
 \qquad \text{(see \eqref{eq:neighborhoods theta})} \nonumber\\
 & = \sum_{m,n=1}^M b_{m,n} \int_{\I^2} \phi_m(t) \phi_n(s)\, h(\der t,\der s)
 \qquad \text{(see \eqref{eq:QFarAway})} \nonumber\\
 & = \l\langle b , \int_{\I^2} \Phi(\theta) \, h(\der \theta) \r\rangle
 \qquad \text{(see \eqref{eq:meas model})}
 \nonumber\\
 & \le  \| b\|_{\mathrm{F}} \cdot \l\| \int_{\I^2} \Phi(\theta) \,h(\der \theta) \r\|_{\mathrm{F}}
 \qquad \text{(Cauchy-Schwarz inequality)}
 \nonumber\\
 & \le \|b\|_{\F} \cdot 2\d', \qquad \text{(see \eqref{eq:feasUsed})}
\end{align}
which completes the proof of Lemma~\ref{lem:dual cert noise}. Above, we note that the matrix $b\in \mathbb{R}^{M\times M}$ is formed by the coefficients $\{b_{m,n}\}_{m,n=1}^M$.

\section{Proof of Lemma \ref{lem:dual cert noise 2}}\label{sec:proof dual cert noise 2}

For notational convenience, throughout we model $x_{K,\varepsilon}$  by~\eqref{eq:atomic}, with $x$ therein replaced with $x_{K,\varepsilon}$. 
The existence of the dual certificate $Q^0$ allows us to write that 
\begin{align}
    & \sum_{k=1}^K \l| \int_{\theta_{k,\varepsilon}} h(\der \theta) \r|  \nonumber\\
    & = \sum_{k=1}^K \int_{\theta_{k,\varepsilon}} s_k \, h(\der \theta)
    \qquad \l(s_k := \mathrm{sign}\l( \int_{\theta_{k,\varepsilon}} h(\der\theta)\r)\r) \nonumber\\
    & = \sum_{k=1}^K \int_{\theta_{k,\varepsilon}} (s_k-Q^0(\theta)) \, h(\der \theta)+ \sum_{k=1}^K \int_{\theta_{k,\varepsilon}} Q^0(\theta) \, h(\der \theta) \nonumber\\
    & = \sum_{k=1}^K \int_{\theta_{k,\varepsilon}} (s_k-Q^0(\theta)) \, h(\der \theta) + \int_{\I^2} Q^0(\theta) \, h(\der \theta) - \int_{\Theta^C_\varepsilon} Q^0(\theta)\, h(\der \theta)
    \qquad \text{(see \eqref{eq:neighborhoods theta})} \nonumber\\
    & = \sum_{s_k = 1} \int_{\theta_{k,\varepsilon}} (1-Q^0(\theta)) \, h(\der \theta) + \sum_{s_k=-1} \int_{\theta_{k,\varepsilon}} (-1-Q^0(\theta)) \, h(\der \theta) \nonumber\\
    & \qquad + \int_{\I^2} Q^0(\theta) \, h(\der \theta) - \int_{\Theta^C_\varepsilon} Q^0(\theta)\, h(\der \theta) \nonumber\\
    & = \sum_{s_k = 1} \int_{\theta_{k,\varepsilon}} (1-Q^0(\theta)) \, h(\der \theta) + \sum_{s_k=-1} \int_{\theta_{k,\varepsilon}\backslash \theta_k} (-1-Q^0(\theta)) \, h(\der \theta)\nonumber\\
    & \qquad +  \sum_{s_k=-1} \int_{\theta_k} (-1+\alpha -Q^0(\theta)) \, h(\der \theta)
    -\alpha  \sum_{s_k=-1} \int_{\theta_k}  h(\der \theta)
    \nonumber\\
    & \qquad + \int_{\I^2} Q^0(\theta) \, h(\der \theta) - \int_{\Theta^C_\varepsilon} Q^0(\theta)\, h(\der \theta). 
\end{align}
Above, we separated the impulses based on the sign of the error near the impulses, and we also singled out the errors at the impulse locations corresponding to the negative sign. In view of~\eqref{eq:q0}, we can bound the last line above by 
\begin{align}
    &   -\alpha  \sum_{s_k=-1} \int_{\theta_k}  h(\der \theta) + \int_{\I^2} Q^0(\theta) \, h(\der \theta) + \int_{\Theta^C_\varepsilon} h(\der \theta)
    \qquad \text{(see \eqref{eq:q0})} \nonumber\\
    &  \le -\alpha  \sum_{s_k=-1} \int_{\theta_k}  h(\der \theta) + \int_{\I^2} Q^0(\theta) \, h(\der \theta) + 2\overline{g}^{-1} \|b\|_\F \d' \qquad \text{(see Lemma~\ref{lem:dual cert noise})} \nonumber\\
    & \le \alpha \sum_{s_k=-1} \int_{\theta_k} x_{K,\varepsilon}(\der \theta) + \int_{\I^2} Q^0(\theta) \, h(\der \theta) + 2\overline{g}^{-1} \|b\|_\F \d'
    \qquad (h = \widehat{x}- x_{K,\varepsilon} \text{ and } \widehat{x} \ge 0)
    \nonumber\\
    & = \alpha \sum_{s_k=-1} a_k + \int_{\I^2} Q^0(\theta)\, h(\der \theta) + 2\overline{g}^{-1} \|b\|_\F \d'
    \qquad \text{(see \eqref{eq:atomic})} \nonumber\\
    & \le \alpha \|x_{K,\varepsilon}\|_{\mathrm{TV}} + \int_{\I^2} Q^0(\theta)\, h(\der \theta)+ 2\overline{g}^{-1} \|b\|_\F \d' \nonumber\\
    & = \alpha \|x_{K,\varepsilon}\|_{\mathrm{TV}} + \l\langle  b^0, \int_{\I^2} \Phi(\theta)\, h(\der \theta)\r \rangle + 2\overline{g}^{-1} \|b\|_\F \d'
    \qquad \text{(see \eqref{eq:meas model})}
    \nonumber\\
    & \le \alpha \|x_{K,\varepsilon}\|_{\mathrm{TV}} + \|b^0\|_{\F} \cdot 2\d' + 2\overline{g}^{-1} \|b\|_\F \d',
    \qquad \text{(Cauchy-Schwarz and \eqref{eq:feasUsed})}
\end{align}
which completes the proof of Lemma~\ref{lem:dual cert noise 2}. Above, the matrix $b^0\in \mathbb{R}^{M\times M}$ is formed by the coefficients $\{b^0_{m,n}\}_{m,n=1}^M$.

\section{Proof of Proposition \ref{prop:existence of 1st dual} \label{sec:Proof-of-Proposition existence of 1st dual}}

The high-level strategy of the proof is again to construct the desired polynomial $Q(\theta)$ in $\theta=(t,s)$ by  combining a number of univariate polynomials in $t$ and $s$. Each of these univariate polynomials is built using a  1-D version of Proposition~\ref{prop:existence of 1st dual}, which is summarized below for the convenience of the reader, see  \cite[Proposition \editr{19}]{eftekhari2018sparse}.\footnote{\editr{{In the proof of Proposition~19} in~\cite{eftekhari2018sparse} and with the notation therein,  $F$ must be such that $\{F\}\cup\{\phi_j\}_{j=1}^m$  form a $\mathcal{C}^*$-system, but is otherwise arbitrary. Our Proposition~\ref{prop:prop 11 rep} thus follows from \cite[Proposition 19]{eftekhari2018sparse}, after \circled{1}~recalling the first property of $\mathcal{C}^*$-systems listed in Remark~\ref{rem:propsTStarSys}, and \circled{2}~noting that the sum in the definition of $\dot{q}$ in the proof of Proposition~19 is a continuous function of $t$ because of the continuity of the functions $\{\phi_j\}_{j=1}^m$.  } } 
\begin{prop}[\editr{Univariate polynomial of a $\mathcal{C}^*$-system}]
\label{prop:prop 11 rep}Consider  a finite set $T' \subset\I$ of size \editr{no larger than $K$}. For $M\ge2\editr{K}+2$, suppose that $\{\phi_{m}\}_{m=1}^M$ form
a $\mathcal{C}$-system on $\I$. Consider also $F':\mathbb{R}\rightarrow\mathbb{R}$ and suppose that $\{F'\}\cup\{\phi_{m}\}_{m=1}^M$ form \editr{a} $\mathcal{C}^*_{K,\varepsilon}$-system on $\I$. Then there exist real
coefficients $\{b_{m}\}_{m=1}^M$ and a \editr{continuous} polynomial $q_{T'}=\sum_{m=1}^M b_{m}\phi_{m}$
such that $q_{T'}\ge F'$ with equality holding on $T'$. 
\end{prop}
Let us now use Proposition \ref{prop:prop 11 rep} to complete the proof of Proposition \ref{prop:existence of 1st dual}. Fix an index set $\Omega\subseteq[K]$. By assumption, $\{F_{T_{\Omega}}\}\cup\{\phi_{m}\}_{m=1}^M$ form \editr{a} $\mathcal{C}^*_{K,\varepsilon}$-system on $\I$. Therefore, by Proposition~\ref{prop:prop 11 rep}, there exists a polynomial $q_{T_{\Omega}}$ such that 
\begin{equation}
q_{T_{\Omega}}\ge F_{T_{\Omega}},
\label{eq:g_T_omega}
\end{equation}
 with  equality holding on \editr{$T_\Omega$}. Likewise, by assumption, $\{F_{S_{[K]\backslash \Omega}}\} \cup \{\phi_m\}_{m=1}^M$   form \editr{a} $\mathcal{C}^*_{K,\varepsilon}$-system on $\I$ and therefore there exists a polynomial $q_{S_{[K]\backslash\Omega}}$ such that 
\begin{equation} 
q_{S_{[K]\backslash\Omega}}\ge F_{S_{[K]\backslash\Omega}},
\label{eq:g_S_omega}
\end{equation} 
with  equality holding on $S_{[K]\backslash \Omega}$.

 As in the proof of Proposition~\ref{prop:existence of dual cert}, consider the polynomial 
\begin{equation}
Q(\theta)=Q(t,s)=\sum_{\Omega\subseteq[K]}q_{T_{\Omega}}(t)\cdot q_{S_{[K]\backslash\Omega}}(s),\label{eq:poly sum}
\end{equation}
where the sum is over all subsets of $[K]$. We next show that $Q$  is the desired dual certificate, \editr{prescribed in Lemma~\ref{prop:existence of dual cert}}. Recall the neighbourhoods defined in~(\ref{eq:neigh t},\ref{eq:neighborhoods theta}).  Fix \editr{an index set $\Omega\subseteq [K]$ and} $k\in[K]$. Assume that $k\in \Omega$. 
For \editr{every} $\theta\in\theta_{k,\varepsilon}= t_{k,\varepsilon}\times s_{k,\varepsilon}$,
\editr{it then holds that}
\begin{align*}
q_{T_{\Omega}}(t)\cdot q_{S_{[K]\backslash\Omega}}(s)+q_{T_{[K]\backslash\Omega}}(t)\cdot q_{S_{\Omega}}(s) & \ge F_{T_{\Omega}}(t)\cdot F_{S_{[K]\backslash\Omega}}(s)+F_{T_{[K]\backslash\Omega}}(t)\cdot F_{S_{\Omega}}(s)
\qquad \mbox{(see~(\ref{eq:g_T_omega},\ref{eq:g_S_omega}))}
\\
 & \ge 0\cdot\pline+\pline\cdot 0=0,
 \qquad \editr{\text{(see \eqref{eq:P_T'})}}
\end{align*}
and the equality in the first line above holds \editr{at least} at $\theta_{k}=(t_{k},s_{k})$. 
\editr{In fact,} the above statement holds also when $k\in[K]\backslash\Omega$.
By summing up over all pairs $(\Omega,[K]\backslash\Omega)$ and \editr{then} applying
the above inequality, we \editr{arrive at}
\begin{align}
Q(\theta) & =\sum_{\Omega\subseteq[K]}q_{T_{\Omega}}(t)\cdot q_{S_{[K]\backslash\Omega}}(s)
\editr{ \ge  0},
\end{align}
which, \editr{to reiterate}, holds for \editr{every} $\theta\in\theta_{k,\varepsilon}$ and with equality at $\theta_k$. The above bound is independent of $k$ and we therefore conclude that 
\begin{equation}
Q(\theta) \ge 0, \qquad \theta \in \Theta_\varepsilon,
\label{eq:Q bnd 1}
\end{equation}
with equality holding at $\Theta$, see~\eqref{eq:neighborhoods theta}.

On the other hand, consider $\theta\in\Theta_{\varepsilon}^{C}$, \editr{i.e., $\theta$} does not belong to \editr{any of the neighbourhoods} $\{\theta_{k,\varepsilon}\}_{k=1}^K$.  \editr{We consider two cases below:}
\begin{enumerate}[leftmargin=*]
\item The first possibility is that 
$$
\theta=(t,s)\in T_{\varepsilon}^{C}\times S_{\varepsilon}^{C}\subseteq\Theta_{\varepsilon}^{C}
.
$$
%i.e., that $t\in T_{\varepsilon}^{C}$ and $s\in S_{\varepsilon}^{C}$. 
In this case, \editr{in view of~(\ref{eq:g_T_omega},\ref{eq:g_S_omega}), it holds} that 
\begin{align}
q_{T_{\Omega}}(t)\cdot q_{S_{[K]\backslash\Omega}}(s) & \ge F_{T_{\Omega}}(t)\cdot F_{S_{[K]\backslash\Omega}}\edita{(s)} \ge\pline,
\qquad \editr{\text{(see \eqref{eq:P_T'})}}
\end{align}
for every index set $\Omega \subseteq [K]$. \editr{By summing up over all index sets, it immediately follows that}
\begin{align}
Q(\theta)& =\sum_{\Omega\subseteq[K]}q_{T_{\Omega}}(t)\cdot q_{S_{[K]\backslash\Omega}}(s)
 \nonumber\\
& \ge 2^{K},\qquad \theta\in T_{\varepsilon}^{C}\times S_{\varepsilon}^{C}.\label{eq:Q bnd 2}
\end{align}
\item The second possibility is that 
$
\theta \in \Theta_{\varepsilon}^{C}\backslash\left(T_{\varepsilon}^{C}\times S_{\varepsilon}^{C}\right).
$
In this case, there exists \editr{a distinct pair} $k,l\in [K]$ such that  $\theta=(t,s)\in t_{k,\varepsilon}\times s_{l,\varepsilon}$ and an index set $\Omega_{0}\subset[K]$
such that $t_{k}\in[K]\backslash\Omega_{0}$ and $s_{l}\in\Omega_{0}$.
It follows that 
\begin{align}
q_{T_{\Omega_{0}}}(t)\cdot q_{S_{[K]\backslash\Omega_{0}}}(s)& \ge F_{T_{\Omega_{0}}}(t)\cdot F_{S_{[K]\backslash\Omega_{0}}}(s) 
 \ge\pline.
 \qquad \editr{\text{(see \eqref{eq:P_T'})}}
\end{align}
There are in fact $2^{K-2}$ such subsets of $[K]$ and we conclude that 
\begin{align} 
Q(\theta) & =\sum_{\Omega\subseteq[K]}q_{T_{\Omega}}(t)\cdot q_{S_{[K]\backslash\Omega}}(s)
\nonumber\\
& \ge2^{K-2},\qquad\theta\in\Theta_{\varepsilon}^{C}\backslash\left(T_{\varepsilon}^{C}\times S_{\varepsilon}^{C}\right).\label{eq:Q bnd 3}
\end{align}
\end{enumerate}
By combining (\ref{eq:Q bnd 2}) and (\ref{eq:Q bnd 3}), we \editr{reach} 
that 
\begin{equation}
Q(\theta)\ge2^{K-2}=:\oline,\qquad\theta\in\Theta_\varepsilon^{C}.\label{eq:Q bnd 4}
\end{equation}
\editr{Lastly,} combining (\ref{eq:Q bnd 1}) and (\ref{eq:Q bnd 4}) completes the
proof of  \editr{Proposition~\ref{prop:existence of 1st dual}.}

\section{Proof of Proposition \ref{prop:existence of 2nd dual} \label{sec:Proof-of-Proposition existence of 2nd dual}}

{The proof is based on the same principles \editr{that appeared in} Appendix~\ref{sec:Proof-of-Proposition existence of 1st dual}.} \editr{Let us fix an arbitrary sign pattern $\pi \in \{\pm 1\}_{k=1}^K$.} 
\editr{For every $k\in [K]$ and by assumption, $\{F_{s_k}^+\}\cup\{\phi_m\}_{m=1}^M$ form a $\mathcal{C}^*_{K,\varepsilon}$-system on $\I$, see~\eqref{eq:GkTnS}. Therefore, by Proposition~\ref{prop:prop 11 rep}, there exists for every $k\in [K]$ a polynomial $q^{\pi_k}_{s_k}$ such that 
\begin{equation}
q^{\pi_k}_{s_k}(s)\ge 
\begin{cases}
F_{s_k}^+(s) & \text{when } \pi_k = 1\\
\varepsilon F_{s_k}^+ (s) & \text{when } \pi_k = -1,
\end{cases}
\label{eq:gkGkS}
\end{equation}
for every $s\in \I$, 
 with  equality holding on $S=\{s_i\}_{i=1}^K$. When the sign pattern $\pi$ contains at least one negative sign, we define for future use the normalized maximum
 \begin{align}
      q_{\max}^\pi (\varepsilon) := \varepsilon^{-1} \max_{\pi_k=-1}\max_{s\in \I} \,\, q_{s_k}^{\pi_k}(s),
      \label{eq:qMax}
 \end{align}}\editr{where the inner maximum above is indeed achieved in view of the continuity of $q_{s_k}^{\pi_k}$ and the compactness of $\I$, see Proposition~\ref{prop:prop 11 rep}. Note that $q_{\max}^\pi(\varepsilon)\ge 1$ for every $\varepsilon>0$, see~(\ref{eq:GkTnS},\ref{eq:gkGkS}).  When $\varepsilon=0$, we  choose the trivial polynomial $q_{s_k}^{\pi_k}=\varepsilon=0$ for every $k$ such that $\pi_k=-1$, and thus record that 
 \begin{equation}
     q_{\max}^\pi (0) = 1.\label{eq:limitCase}
 \end{equation}}\editr{Likewise,} 
for every $k\in [K]$,  $\{F_{t_k}^{\pi_k}\}\cup\{\phi_m\}_{m=1}^M$ form a $\mathcal{C}^*_{K,\varepsilon}$-system on $\I$ \editr{by assumption}, see \eqref{eq:GkTnS}.  Therefore,
by Proposition~\ref{prop:prop 11 rep},  there exists for every $k\in [K]$  a polynomial
$q^{\pi_k}_{t_k} $ \editr{such that} 
\editr{
\begin{equation}
q_{t_k}^{\pi_k} (t)\ge 
\begin{cases}
F_{t_k}^{\pi_k}(t) & \text{when } \pi_k = 1 \\
\\
\frac{F_{t_k}^{\pi_k} (t)}{\varepsilon q_{\max}^\pi(\varepsilon)}  
& \text{when } \pi_k = -1,
\end{cases}
\label{eq:gkGkT}
\end{equation} 
for every $t\in \I$,} 
with  equality holding on \editr{$T=\{t_i\}_{i=1}^K$}. 
 \editr{In view of~(\ref{eq:gkGkS},\ref{eq:gkGkT}), and after recalling  the definitions of $F^+_{s_k}$ and $F^{\pm}_{t_k}$ from~\eqref{eq:GkTnS}, we observe that the product $q_{t_k}^{\pi_k}(t) q_{s_k}^{\pi_k}(s)$ satisfies
\begin{align}
  q_{t_k}^{\pi_k}(t)q_{s_k}^{\pi_k}(s) & 
  \ge 
  \begin{cases}
  \pi_k & \text{when }\theta\in \theta_{k,\varepsilon} \\
  - \frac{1}{q^\pi_{\max}(\varepsilon)} & \text{when }\theta = \theta_k \text{ and } \pi_k = -1\\
  -1 & \text{when }t\in t_{k,\varepsilon} \text{ and } \pi_k = -1\\
  0 & \text{elsewhere in } \I^2,
  \end{cases}
  \label{eq:onePairProd}
 \qquad 
\end{align}
with equality holding at least on $\Theta$. The third case above indicates that there is a stripe in $\I^2$ over which the product $q_{t_k}^{\pi_k}q_{s_k}^{\pi_k}$ is bounded below by $-1$.} Let us \editr{now} consider the polynomial 
\[
Q^{\pi}(\theta)=Q^\pi(t,s):=\sum_{k=1}^K \editr{q^{\pi_k}_{t_k}(t)\cdot q^{\pi_k}_{s_k}(s)}.
\]
\editr{We next establish that, for the appropriate choice of the sign pattern $\pi$, $Q^\pi$
% \tb{What do you mean by that?} 
is indeed the desired dual certificate prescribed in Lemma~\ref{lem:dual cert noise 2}. Recall that impulses are $\varepsilon$-separated  and that, in particular,~\eqref{eq:sep 1d} holds for the choice of $\nu=\varepsilon$ therein.  Then, we invoke~\eqref{eq:onePairProd}  to write5} that 
\begin{align}
Q^{\pi}(\theta) & = \sum_{k=1}^K 
\editr{q_{t_k}^{\pi_k}(t)\cdot q^{\pi_k}_{s_k}(s)} \nonumber\\
 & \ge\begin{cases}
\pi_k & \text{when \editr{there exists }} \editr{k\in [K]} \editr{\text{ such that }}\theta\in\theta_{k,\varepsilon}\\
 \editr{-\frac{1}{q_{\max}^\pi (\varepsilon)}}& \editr{\text{when } \text{\editr{there exists }} \editr{k\in [K]} \editr{\text{ such that }}\theta = \theta_k \text{ and } \pi_k = -1}\\
\editr{-1} & \text{elsewhere \editr{in} }\I^2,
\end{cases}
\label{eq:Qpi is large}
\end{align}
and the equality holds \editr{at least} on $\Theta$.
% Let $\pi^0$ be the sign pattern of $\{\int_{\theta_{k,\varepsilon}}h(d\theta)\}_{k=1}^K$. Then,~\eqref{eq:Qpi is large} implies that $Q^ 0 := Q^{\pi_0}\ge G^0$, see~\eqref{eq:q0}. 
\edita{Finally, since our choice of the sign pattern $\pi$ in the beginning of the proof was arbitrary, the existence of the dual certificate in Lemma~\ref{lem:dual cert noise 2} is also guaranteed, even though the sign pattern of the error near the point sources is unknown to us a priori.} \editr{More specifically, let $\pi_*$ denote the sign pattern specified by the error measure $h$ in~\eqref{eq:q0}. Then the dual certificate in Lemma~\ref{lem:dual cert noise 2} exists with 
\begin{align}
\alpha(\varepsilon) = 1- 1/q_{\max}^{\pi_*}(\varepsilon).
\label{eq:alphaDefined}
\end{align}
In particular, $\alpha(0)=0$ by~\eqref{eq:limitCase}.}  
This completes the proof of Proposition~\ref{prop:existence of 2nd dual}. 

\bibliographystyle{unsrt}
\bibliography{References}

\end{document}